\documentclass[prl,twocolumn,nofootinbib]{revtex4}
\usepackage[utf8]{inputenc}
\usepackage{amsmath}
\usepackage{amsfonts}
\usepackage{amssymb}
\usepackage{graphicx}
\usepackage{grffile}
\usepackage{color}
\usepackage{textgreek}
\usepackage{hyperref}
\usepackage{multirow}
\usepackage[normalem]{ulem}
\begin{document}
\title{Soft-Pinning: Experimental Validation of Static Correlations in Supercooled Molecular Glass-forming Liquids. }
\author{Rajsekhar Das$^{1,2}$}
\author{Bhanu Prasad Bhowmik$^{1,3}$}
\author{Anand B. Puthirath$^{1,4}$}
\author{Tharangattu N. Narayanan$^{1}$}
\author{Smarajit Karmakar$^{1}$}
\email{smarajit@tifrh.res.in}
\affiliation{
$^1$TIFR Center for Interdisciplinary Science, Tata Institute of Fundamental Research,36/P Gopanpally Village, Serilingampally Mandal, RR District\\
$^2$ Department of Chemistry, University of Texas at Austin, Austin, Texas 78712, USA\\
$^3$ Weizmann Institute of Science, Rehovot, Israel\\
$^4$ Department of Materials Science and Nano Engineering, Rice University, 6100 Main Street Houston, TX 77005, USA 
}

\begin{abstract}
Enormous enhancement in the viscosity of a liquid near its glass transition is generally connected to the growing many-body static correlations near the transition, often coined as ‘amorphous ordering’. Estimating the length scales of such correlations in different glass-forming liquids is highly important to unravel the physics of glass formation. 
Experiments on molecular glass-forming liquids become pivotal in this scenario as the viscosity grows several folds ($\sim 10^{14}$), 
simulations or colloidal glass experiments fail to access the long-time scales required. Here we design an experiment to extract the static length scales in molecular liquids using dilute amounts of another large molecule as a pinning site. Results from 
dielectric relaxation experiments on supercooled glycerol with different pinning concentrations of sorbitol and the simulations 
on a few model glass-forming liquids with pinning sites indicates the robustness of the proposed method, opening a plethora 
of opportunity to study the physics of other glass-forming liquids. 
\end{abstract}
\maketitle 
\section{Introduction}
One of the most puzzling states of matter is undoubtedly the glassy state. Glasses are ubiquitous in
nature, yet the physics of glass formation remained unsolved even after many decades of research efforts\cite{Angell1924,Debenedetti2001,Parisi2010,Biroli2011}.
Understanding the dynamical and mechanical properties of glass-forming liquids and solids are important as similar
dynamical and mechanical behaviour can be found in various complex systems ranging from colloidal assemblies \cite{Chaikin1982,vanderScheer2017,Cipelletti2018,Ranjini2015}
to biologically relevant systems like cell membrane, cell migration and bio-preservation\cite{schoetz2013glassy,hardin2013glassy,bi2014energy,sokolov1999glassy,newman1993role}. A detailed understanding
of the fundamental properties of glass-forming systems will have far-reaching implications both in fundamental science
and in industrial applications. One of the major challenges in the field of glass transition is to understand the microscopic origin of the dramatic growth
of viscosity of the glass-forming liquid upon a relatively small change in temperature while approaching the laboratory glass
transition ($T_G$), defined empirically as the temperature where relaxation time reaches $100sec$.
Viscosity in some of these systems can change by $13-14$ orders of magnitude upon $50^{o}-100^{o}$
a shift in temperature from its $T_G$. Another sought-after question in this field is whether glass transition is a true thermodynamic
phase transition or a purely dynamical cross-over. There are a few existing theories that support both of these ideas. On the one
hand, there is Random First-Order Transition (RFOT) theory, introduced by Kirkpatrick, Thirumalai, and Wolynes~\cite{RFOT1, RFOT2}
and later reformulated by Bouchaud and Biroli~\cite{bouchaudBiroliPTS}, which predicts the glass transition to be a thermodynamic
phase transition, and glassy state is a real thermodynamic state. On the other hand, there are kinetically constrained
models, and mode-coupling theory, which describes glass transition as a dynamic phenomenon \cite{InhomoMCT, Garrahan2011}.      

\noindent Numerous works in the last few decades on the glass transition suggest that the rapid growth of viscosity can be attributed to a growing static length scale \cite{smarajitReview,Biroli2008,bouchaudBiroliPTS}. Yet another growing length-scale that can be identified is the dynamic heterogeneity length-scale
~\cite{Sastry2010,Donati1999,Berthier1797,KDS2009} which physically
means the typical size of dynamically correlated regions in the system and are often quantified via four-point susceptibility, $\chi_4(t)$, defined in the Method section. On the other hand, the static length-scale is intrinsically related to structure and is commonly believed to refer to the growth of the so-called 'amorphous order.' One can consider the growth of amorphous order in supercooled liquid as the growth of a region where particles rearrange cooperatively. The size of the region is directly related to the length-scale. Although there are many existing methods
to extract the desired static length scale, most of them are not useful for experiments on molecular glass-forming liquids. In a recent work ~\cite{RajsekharSoftMatter}, it has been shown that it is possible
to extract static length-scale in a glass-forming liquid by measuring the response of the system in the presence of a small concentration of solute particles, which have a smaller diffusion constant than the constituent molecules or particles of the supercooled liquid medium. The idea mainly originated from the previous works on the effects of random pinning sites on the dynamics of the supercooled liquids~\cite{SaurishScientificReport, RajsekharJCPPinning}. As random pinning in a molecular liquid will be experimentally challenging, the idea of ``soft'' pinning sites was
proposed.  It was found that impurity particles with significantly smaller diffusion constants will behave like a pinning site for the particles
of the liquid over a timescale comparable to the relaxation time of the liquid.
 
\begin{figure*}[!ht]
\begin{center}
\vskip-0.5in
\includegraphics[width = 0.95\textwidth]{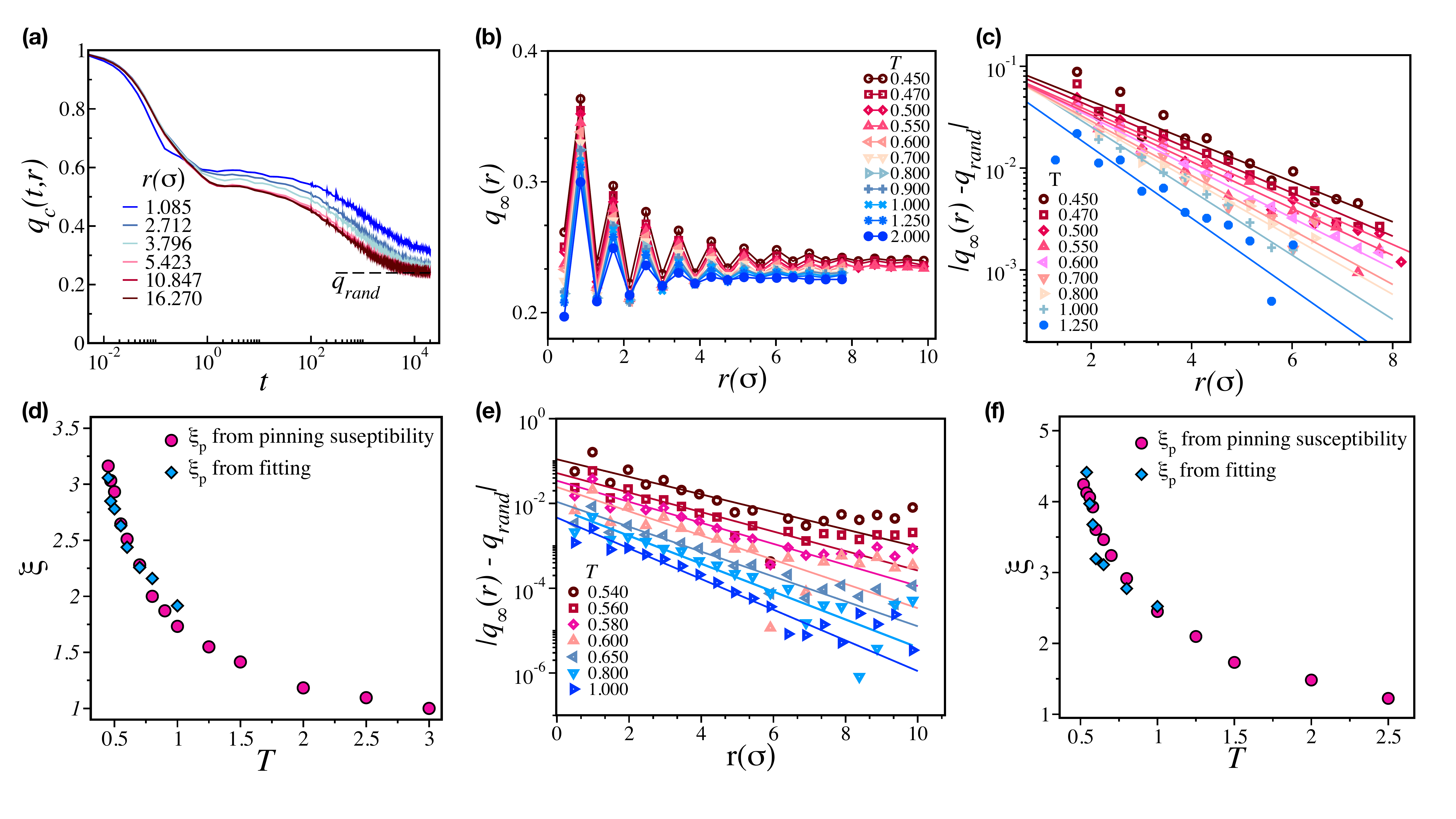}
\caption{\textbf{Estimation of length scales via Random Pinning.} Panel  (a) shows the time dependence of configuraional overlap $q_c(t,r)$ at $T = 0.600$ for the \textbf{2dR10} model. The dashed line correspond to the bulk value $q_{rand}$. Panel (b) shows the variations of $q_{\infty}$ as a function of $r$ for \textbf{2dmKA} model. In panels (c) and (d) $|q - q_{rand}|$ is plotted as a function of $r$ at all the studied temperatures for \textbf{2dmKA} and \textbf{2dR10} models respectively. Solid lines corresponds to the fit to the Eqn.~\eqref{FittingEq} to obtained the correlation lengths. In panels (d) and (f) we compare the length scales obtained from the fitting with that obtained using pinning susceptibility(see text for details) for \textbf{2dmKA} and \textbf{2dR10} models respectively.  The value of $\xi$ increases with decreasing temperatures.} 
\vskip -0.5cm
\label{qctrRPLengthCalc}
\end{center}
\end{figure*}
\noindent An elegant yet simple scaling theory can be developed if one assumes that each pinning site slows down the relaxation process in its
the immediate neighbourhood of liner size, $\xi_s$, where $\xi_s$ is the static correlation length,
then the overall relaxation time of the entire system will be slower with increasing concentration($c$) of the pinning sites.
Now, if one assumes that the relaxation time around a pinning site is solely controlled by the underlying static length scale, then one 
expect that the radial dependence of the relaxation time from the pinning site will be a scaling function as given below
\begin{equation}
\tau_\alpha^p(r,T) = \tau_\alpha(T)\mathcal{F}\left[ \frac{r}{\xi_s(T)}\right],
\label{taurScalingAnsatz}
\end{equation}
where $\xi_s$ is the underlying static length scale and $\mathcal{F}(x)$ is a scaling function which goes to $1$ as $x\to\infty$.
Here $\tau_\alpha(T)$ is the bulk relaxation time obtained from the decay of two-point density correlation function, $Q(t)$ (defined
in the Method Section). The validity of the above scaling ansatz (Eq.\ref{taurScalingAnsatz}) for the random pinning scenario
has been presented in the SM (FIG.S2); the good quality of the data collapse using appropriate static correlation length indeed suggests
the robustness of the approximation. We can now write the relaxation time of the entire system in the presence of $c$ fraction of pinning sites
(in the dilute limit) as
\begin{eqnarray}
\tau_\alpha(c,T) &=& \frac{1}{N}\left[cN\rho\int_{0}^{\infty}[\tau_\alpha^p(r,T) - \tau_\alpha(T)] r^2dr + N\tau_\alpha(T)\right]\nonumber\\
&=& \tau_\alpha(T) \left[c\rho\int_{0}^{\infty}\left[\mathcal{F}\left(\frac{r}{\xi_s(T)}\right) - 1\right]r^2dr + 1\right]\nonumber\\
&=& \tau_\alpha(T) \left[ 1 + c\rho\xi_s^3\int_{0}^{\infty}[\mathcal{F}(x) - 1]x^2dx\right] \nonumber
\\
&\simeq& \tau_\alpha(T) (1 + \kappa c\xi_s^3),
\end{eqnarray}
where $\kappa = \rho\int_{0}^{\infty}[\mathcal{F}(x) - 1]x^2dx$ and $\rho$ is the number density of the system. Thus, the scaling relation we arrive at is
\begin{equation}
\frac{\tau_\alpha(c,T)}{ \tau_\alpha(T) } \simeq 1 + \kappa c\xi_s^3 \quad \mbox{or} \quad \log{\left[ \frac{\tau_\alpha(c,T)}{ \tau_\alpha(T) } \right]} \simeq \kappa c\xi_s^3,
\label{tauScalingAnsatz}
\end{equation}
in the small pinning concentration limit. The above equation suggests that the ratio of relaxation times is directly related to the
static length-scale. A careful scaling analysis will lead to the extraction of the same. Note that in Ref.\cite{RajsekharSoftMatter},
detailed mathematical arguments lead to the following scaling relation
\begin{equation}
\ln\left[\frac{\tau_{\alpha}(c,T)}{\tau_{\alpha}(0,T)}\right] = f\left[c\xi_s^3(T)\right],
\label{pinningScaling}
\end{equation}  
which is very similar to Eq.\ref{tauScalingAnsatz} if the scaling function is Taylor expanded to keep the first two terms in the expansion
in the small concentration limit. This argument suggests that if one can show that pinning sites do nucleate a correlated volume of size
$\xi_s^3$, whose relaxation time is larger than the bulk, then the above-mentioned scaling ansatz will hold good, and the same can
be used to compute the underlying static length-scale. Using Eq.\ref{tauScalingAnsatz}, a new susceptibility,
``pinning susceptibility" is introduced in Ref.~\cite{RajsekharSoftMatter} to estimate the static length directly, and it is defined as
\begin{equation}
\chi_p(T,t) = \left.\frac{\partial Q(T,c,t)}{\partial c}\right|_{c=0}
\label{pinningSus}
\end{equation}
where $Q(T,c,t)$ is overlap function (see Method section). 
Note that random pinning has recently been studied extensively to understand the possibility of ideal glass transition in these systems
\cite{Cammarota8850,BiroliJCP,Ozawa6914,SaurishPNAS, SaurishScientificReport, RajsekharJCPPinning}. Using simple calculations and a few assumptions in Ref.~\cite{RajsekharSoftMatter} it is shown that $\chi_p^{max}(T) \sim \xi_s^3(T)$.
The relation states that the peak value of $\chi_{p}(T)$ should give the direct measure of $\xi_p$. Using numerical simulation in
Ref.~\cite{RajsekharSoftMatter}, the validity of this relation has been checked, and it has also been shown that this scaling theory
works if one uses large size particles having significantly lesser diffusion coefficients to act as 'soft pinning' sites. This finding makes
this proposal very attractive to use in experiments, unlike hard pinning, as realizing soft pinning in a natural system will be much easier.
It will then be possible to extract static length scales in any glass-forming system in experiments. In the subsequent parts of this article, we show the validity of some of these assumptions and then extract the static length-scale in both model glass-forming
liquids via computer simulations and in supercooled glycerol via dielectric experiments.

\section{ Results}
To understand the effect of particle pinning on the local dynamics in model glass-forming liquids we follow the method
developed in~\cite{Kob2012} and later used in~\cite{Ganapathi2018,Hu6375}. The idea is to quantify
the effect of pinned particles (small pinning fraction ) on the local static order present in the system. To do so,
we first study the overlap between configurations over time using configurational overlap
$q_c(t)$\cite{Kob2012,Ganapathi2018,Hu6375}. The quantity $q_c(t)$ is of significant importance as
it is suggested to be an order parameter in RFOT~\cite{RFOT1,RFOT2}. For the estimation of $q_c(t)$
we first divide the simulation box into a small grid of $0.5\sigma$, $\sigma$ being the particle diameter so that 
one small box can not accommodate more than one particle at a point in time. $q_c(t)$ for each box is then defined as
\begin{equation}
q_c(t) = \frac{\langle n_i(t)n_i(0)\rangle}{\langle n_i(0)\rangle},
\end{equation}
where $n_i(t) = 1$ if the box is occupied by a particle or else $0$. We chose a pinned particle as
center and then calculate the radially averaged value $q_c(t,r) = \langle q_c(t)\rangle_r$ for all the
boxes between $r$ and $r +dr$. Particles close to the pinned particles will remain highly correlated to
each other. However, as one goes further away from pinned particles, the effect will be diminished and
one will capture the bulk value. Thus the decay of $q_c(t,r)$ will be slower for smaller $r's$. At
long times, $q_c(t\to\infty,r)$ reaches a plateau and oscillates around the bulk value $q_{rand}$,
the occupation probability of the individual box. Fig.~\ref{qctrRPLengthCalc}(a) shows a typical plot
of $q_c(t,r)$ for various values of $r$ for temperatures $T = 0.600$ for \textbf{2dR10} model.
Note that $q_c(t,r)$ capture the basic feature of the glassy liquids -- the two
step relaxation which is prominent in Fig.~\ref{qctrRPLengthCalc} (a) for $T = 0.600$. Similar results
for other temperatures are shown in the SM.
\begin{figure*}[!ht]
\begin{center}
\vskip-0.5in
\includegraphics[width = 0.95\textwidth]{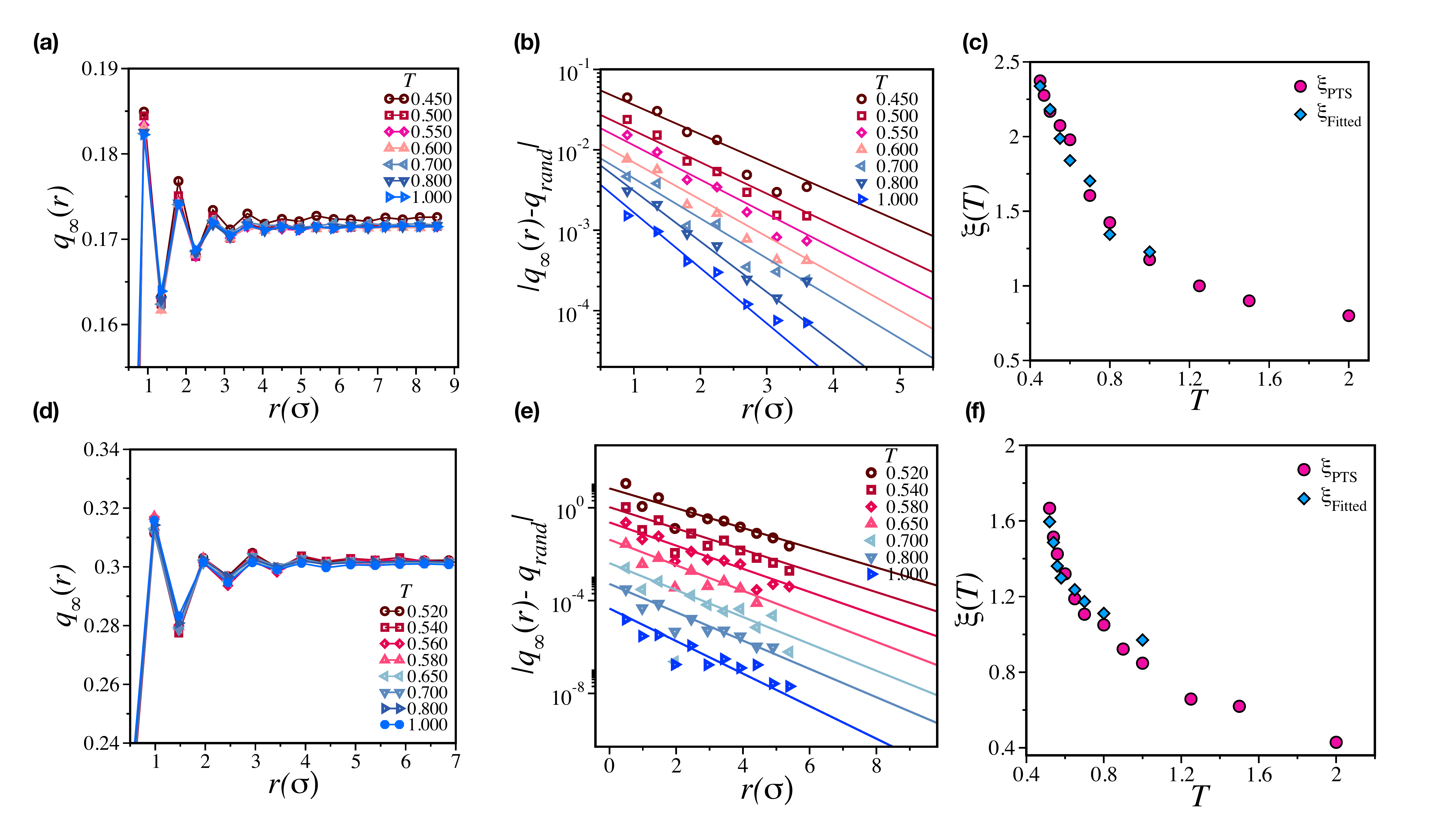}
\caption{\textbf{Estimation of length scales via Soft Pinning.} Panel (a) and (d) shows $q_\infty(r)$ as a function of $r(\sigma)$ for two three dimensional ternary systems -- (a) \textbf{3dKA} and (b) \textbf{3dR10} models respectively. In panels (b) and (e) $|q - q_{rand}|$ is plotted as a function of $r$ at all the studied temperatures for \textbf{3dKA} and \textbf{3dR10} models respectively. Solid lines corresponds to the fit to the Eqn.~\eqref{FittingEq} similar to as shown in Fig.~\ref{qctrRPLengthCalc}. We compare then the obtained length scales with one obtained using PTS (see SM for details) methods in panel (c) and (f) for \textbf{3dKA} and \textbf{3dR10} models respectively.}    
\vskip -0.3in
\label{qinfSP}
\end{center}
\end{figure*}
\subsection{Random pinning }
We estimate the long time value $q_\infty$ of $q_c(t\to \infty,r)$ by averaging the values at the plateau over time. In Fig.~\ref{qctrRPLengthCalc}(b) we
show the oscillation of $q_\infty(r)$ around $q_{rand}$ at various temperatures for \textbf{2dmKA} model. Similar results for \textbf{2dR10} model are
shown in SM. It is important to note that the amplitude of the oscillations around $q_{rand }$ is higher close to the pinned particles, and it gradually
decreases at longer distances. This behavior is similar to that of a pair correlation function and captures the pinned
particles' density fluctuations. The envelope of each curve decays exponentially as a function of $r$ from the pinned center, and decay 
$|q - q_{rand}|$ as a function of
$r$ can easily give us the estimate of the static correlation length. In Fig.~\ref{qctrRPLengthCalc}(c) we depict $|q - q_{rand|}$ as a function of $r$
for \textbf{2dmKA} and Fig.~\ref{qctrRPLengthCalc}(d) shows the results for \textbf{2dR10}. The solid lines are fits to the data using
\vskip -0.2in
\begin{equation}
\label{FittingEq}
|q - q_{rand}| = A\exp(-r/\xi),
\end{equation}
where $A$ and $\xi$ are fitting parameters. 
The length-scale obtained from the fitting increases with decreasing temperatures. We would like to emphasize here that the
length-scale obtained in this way matches well (see Fig.~\ref{qctrRPLengthCalc} (d) and (e) for \textbf{2dmKA} and \textbf{2dR10}
respectively) with the pinning length scale obtained using random pinning scaling analysis described in Ref.~\cite{RajsekharSoftMatter,
RajsekharJCPPinning}. Thus, the method ``Pinning Susceptibility" (for details see Ref.~\cite{RajsekharSoftMatter}) derived within the
the framework of RFOT captures the static length-scale (in this case, the pinning length-scale) of the system unambiguously.
In the subsequent sections, we will show how ``pinning susceptibility" using a ``soft pinning'' site can also be used in a similar manner to
extract the static length-scale of the system. 

\begin{figure*}[!ht]
\begin{center}
\includegraphics[width= 0.95\textwidth]{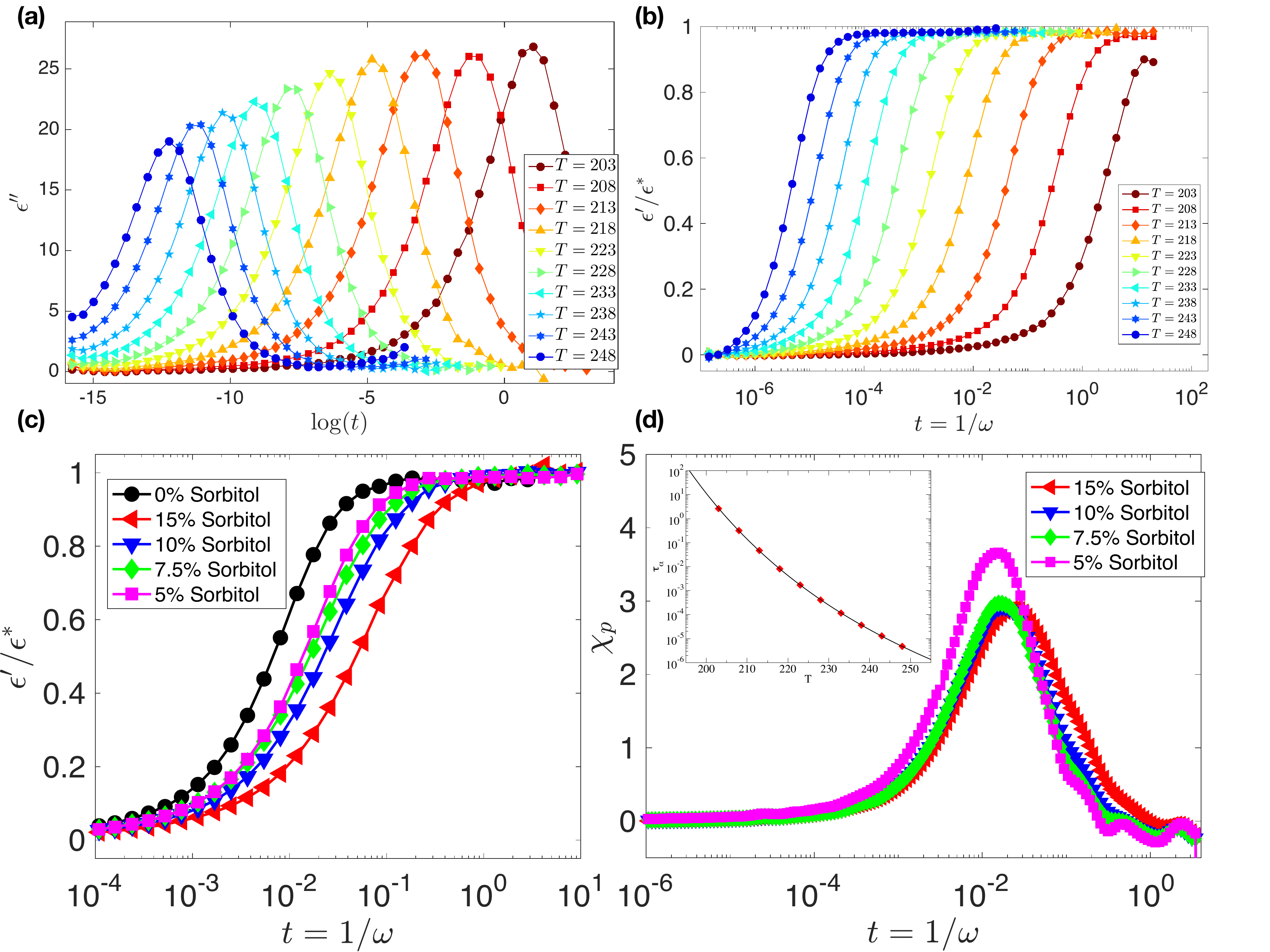}
\caption{ (a) Time evolution of dielectric loss modulus ($\epsilon^{\prime\prime}$) of pure Glycerol. The peak of $\epsilon^{\prime\prime}$ at any particular temperature gives the measure of $\alpha$ relaxation time at that temperature. The time at which $\epsilon^{\prime\prime}$ has maximum value moves to larger time as temperature decreases which manifests the dynamic slowing down. (b) Variation of dielectric storage modulus($\epsilon^{\prime}$) as a function of time (in sec.) for different temperatures for pure Glycerol. Note that $\epsilon^{\prime}$ is re-scaled between $0$ and $1$ using Havriliak-Negami fitting function \cite{HAVRILIAK_NEGAMI}. (c) Variation of dielectric storage modulus($\epsilon^{\prime}$) at $T = 218K$ for various Sorbitol concentrations. The data in the black circle refers to pure Glycerol data. Data in magenta (square), green (diamond), blue (left triangle), red (down triangle) refer for four different Sorbitol concentrations, $5.0\%$, $7.5\%$ $10\%$ and $15\%$ respectively. (d) Shows the pinning susceptibility, $\chi_p$ computed at $T = 218K$ using these
four different concentrations of Sorbitol in Glycerol.  Inset shows the $\tau_{\alpha}$ as a function of temperature. The solid red line stands for VFT fit to extract the Kauzmann temperature, which turns out to be $T_{VFT} \simeq 122K$ in complete agreement with previously reported results.}
\label{exptInitial}
\end{center}
\end{figure*}  
\subsection{Soft pinning }
Recently~\cite{Gokhale2014,Ganapathi2018}, the particle pinning method has been implemented very elegantly in experiments of colloidal glasses.
However, random pinning is still quite challenging to implement in many experiments, particularly in molecular glasses. As discussed in our earlier
work~\cite{RajsekharSoftMatter} and shown in the subsequent paragraphs, the ``soft pinning (SP)" method is a possible way to realize a similar effect as that of random pinning. We devote the next part of our discussion to the effect of ``soft pinning'' on the dynamics of glass-forming liquids both numerically and experimentally.
To achieve the effect of particle pinning, we added a few impurity particles having a larger diameter than the parent liquids particles. Particles with bigger diameters will have smaller diffusion constants, according to the famous Stokes-Einstein relation. They will move much slower than the
solvent particles~\cite{RajsekharSoftMatter} over their timescale of the structural relaxation. Following the definition in Ref.\cite{NagamanasaNaturePhy-Xi-Overlap},
we identify the soft pinning particles by first dividing
the simulation box in a grid of linear size $l = \sigma$. Then we compute $q_c(t)$ for all the impurity particles and those having $q_c(t) > 0.85$ for
$t \sim (5-10 \tau_\alpha)$ are chosen as ``soft pinned" particles. Effectively those particles do not move much over the time scale $5\tau_\alpha$
- $10 \tau_\alpha$ and behaves like randomly pinned particles. Once these soft pinned particles are identified, we can repeat the similar analysis described in the earlier section to obtain the underlying length-scale.

\noindent In Fig.~\ref{qinfSP} (a) and (d) we show $q_\infty(r)$ as a function of $r$ for \textbf{3dKA} model and \textbf{3dR10} model respectively. Details of these
models are given in the Method section. Interestingly, note that oscillation of $q_\infty$ is similar to that observed in the case of random pinning. The oscillation amplitude is substantial for small $r$, and it dies out at larger $r$. Thus, the decay of the peak value of the static correlation as a function of $r$ will capture the static length-scale of the system. In Fig.~\ref{qinfSP}(b) and (e) we show the exponential decay of $|q_\infty - q_{rand}|$ as a function of $r$ the \textbf{3dKA} and \textbf{3dR10} ternary models respectively. The solid lines are the fits to Eqn.~\eqref{FittingEq} to estimate the static length scales. The length scale obtained from the fitting is compared with the length scales obtained from other conventional methods
like the point-to-set (PTS) method (see SM for details of PTS method) in panels (c) and (f) of Fig.~\ref{qinfSP} for \textbf{3dKA} and
\textbf{3dR10} models respectively. The nice agreement between the extracted length-scales from this analysis and previously reported results
demonstrates the robustness of the ``soft pinning'' method in studying the growing static correlations in molecular liquids. This enables
us to test the validity of this method in supercooled glycerol using dielectric spectroscopy experiments, as discussed in the next section.

\vskip +0.1in
\begin{figure*}[!ht]
\begin{center}
\includegraphics[width = 0.98\textwidth]{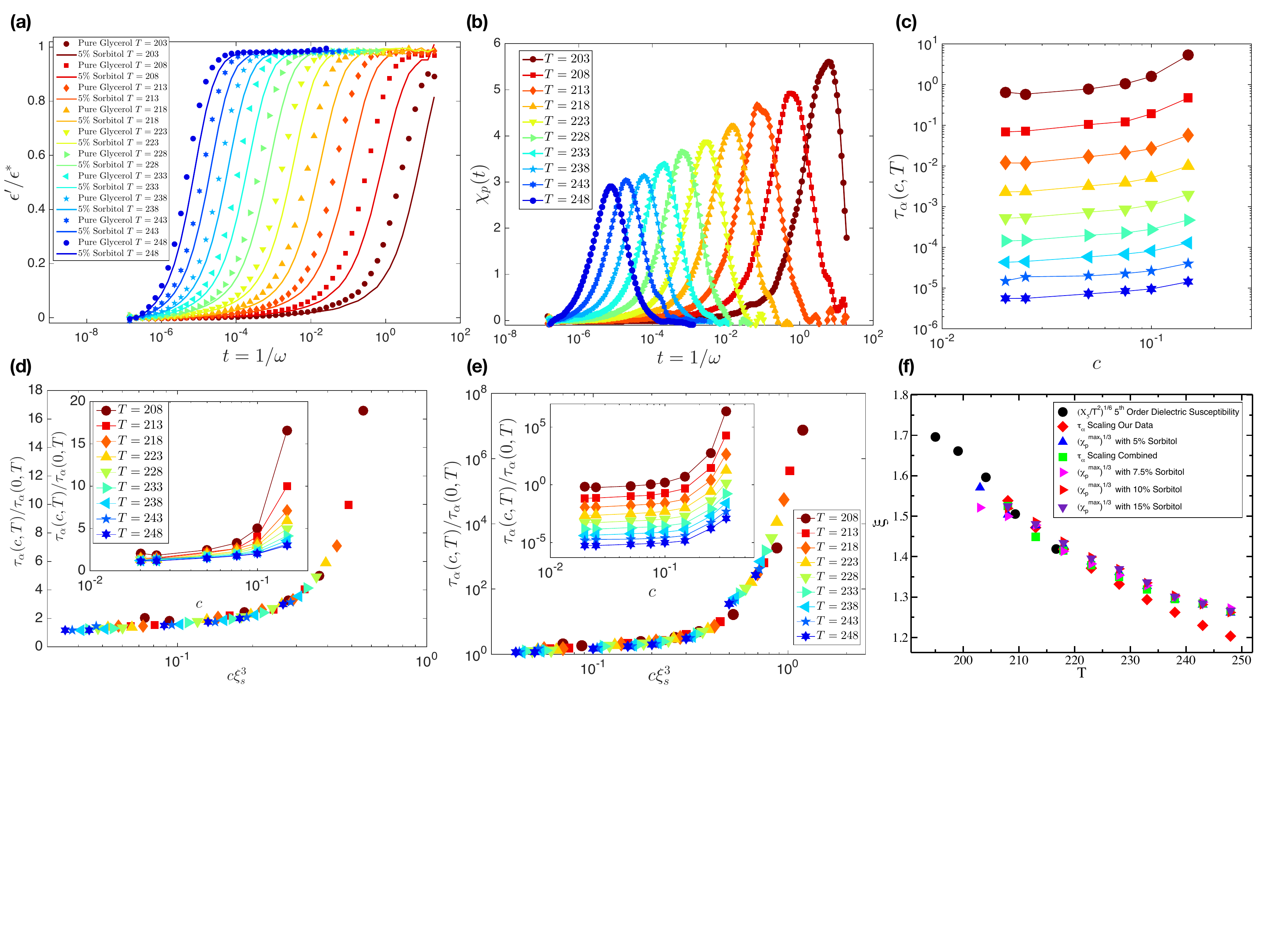}
\vskip -1.4in
\caption{(a)Time dependency of dielectric storage modulus for pure Glycerol and $7.5\%$ Sorbitol mixture. (b) Pinning susceptibility as a function of time.(c) Variation of $\alpha$ relaxation time of Glycerol Sorbitol mixture as a function of Sorbitol concentrations  for different temperature.(d) Data collapse of $\tau_{\alpha}$ using the scaling relation Eq.\ref{pinningScaling}. (e) Data collapse of $\alpha$ relaxation time of Glycerol-Sorbitol mixture. (f) Comparison of static length scale found from the experiment and calculated from the data provided in Ref.~\cite{ Shachi_exptDataPRL, BiroliScienceGlucerol-2}. Note that length
scales obtained from peak values of $\chi_p$ at various sorbitol concentrations gives the same length scale. This surely is a benchmark supporting evidence
of the validity of the scaling assumptions as well as the reliability of the obtained length scale.}
\label{exptGlycerol}
\vskip -0.2in
\end{center}
\end{figure*}	
\subsection{Dielectric Experiments - Glycerol-Sorbitol Mixture }	
We now come to the experimental verification of our ``soft pinning'' method using dielectric spectroscopy measurements on supercooled
glycerol using sorbitol as ``soft pinning'' co-solvent. Here we present the salient features of our experimental results, and the details of
the experimental procedures are described in the Method section. In the Fig.~\ref{exptInitial} (a) and (b), we show variation of the loss modulus
($\epsilon^{\prime\prime}$) and normalized storage modulus ($\epsilon^{\prime}$)of pure Glycerol as a function of time at different temperatures.
The normalization of the $\epsilon^{\prime\prime}$ is done by fitting the data using Havriliak-Negami fitting function \cite{HAVRILIAK_NEGAMI}
and then re-scaling it between $0$ and $1$. The $\alpha$-relaxation time ($\tau_{\alpha}$) of the system is obtained from the peak of
$\epsilon^{\prime\prime}$. The peak position shifts to the higher time with decreasing temperature, which is a clear indication of slower dynamics
at a lower temperature. The temperature dependency of $\tau_{\alpha}$ is shown in the inset of Fig.~\ref{exptInitial} (d). We fit the these data
using well known VFT function $\tau = \tau_0\exp{\left(\frac{A}{T-T_0}\right)}$ and estimated the VFT temperature $T_{VFT} \simeq 122K$
and the calorimetric glass transition temperature of glycerol, $T_G \simeq 195K$. $T_G$ is defined as the temperature at which the relaxation
time becomes $100s$. These results are in close agreement with the value reported in \cite{Duvvuri2004_GlycerolandSorbitol}.
In Fig.~\ref{exptInitial} (c), we show the storage modulus at $T = 218K$ for various concentrations of Sorbitol in Glycerol. The data represented
by black circle refers to pure Glycerol at $T = 218K$. Data in magenta (square), green (diamond), blue (left triangle), red (down triangle) refer
for four different Sorbitol concentrations, $5.0\%$, $7.5\%$ $10\%$ and $15\%$ respectively. Fig.~\ref{exptInitial} (d) presents the pinning
susceptibility, $\chi_p \equiv \partial \epsilon^{\prime}(t)/\partial c$, where $c$ is the concentration of Sorbitol (see Method section).
The four data sets are for the Sorbitol
concentrations of $5.0\%$, $7.5\%$ $10\%$ and $15\%$ respectively. Note that $\chi_p$ shows a nice peak at the typical relaxation time of
the system. The peak height, which is supposed to measure the volume of the static correlation, remains more or less constant according to the scaling theory presented before, with changing concentrations of sorbitol. This confirms that indeed pinning susceptibility
is a robust measure of the growing correlations in molecular glass-forming liquids, as elaborated in detail in the following paragraph.

\begin{figure*}[!ht]
\begin{center}
\includegraphics[width = 0.95\textwidth]{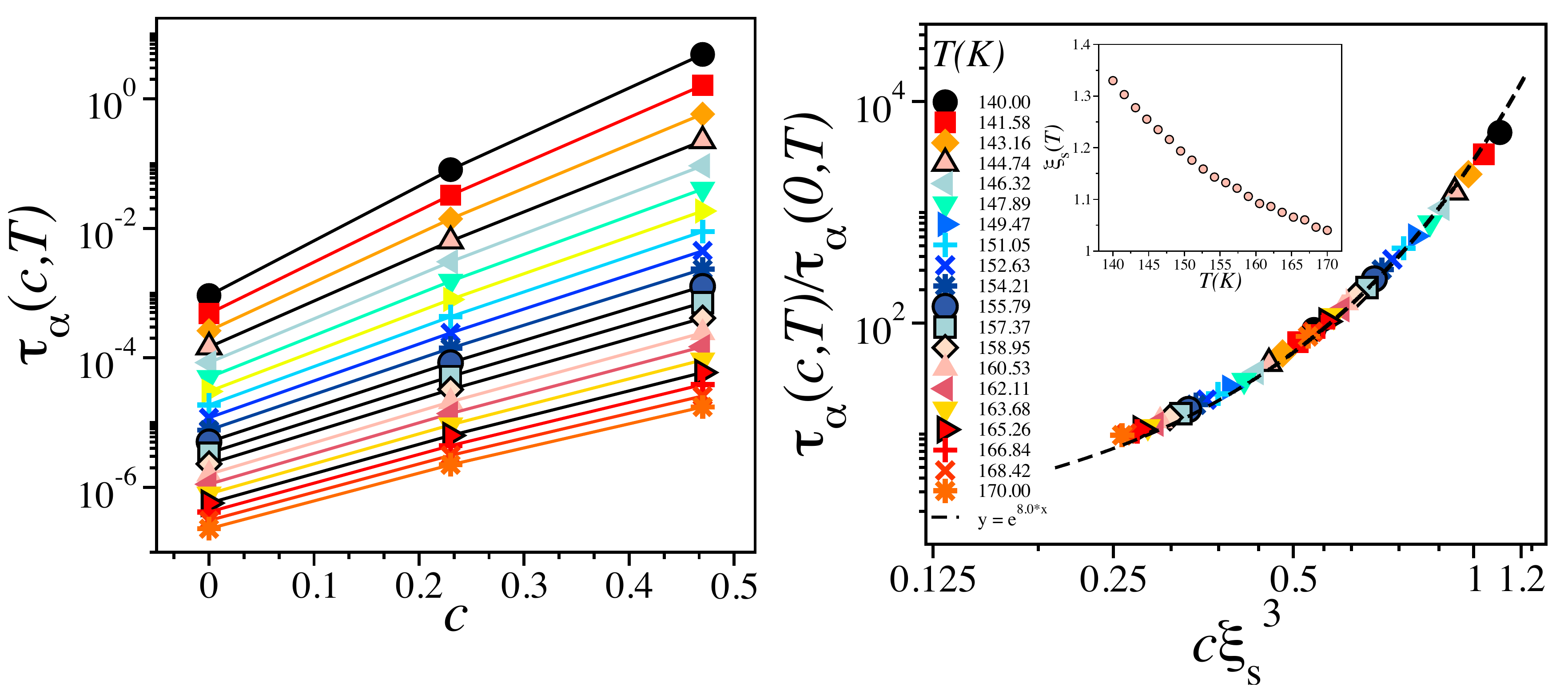}
\vskip -0.1in
\caption{Left panel: The extracted data of Butanol-Hexanol binary glass forming mixture as a function of concentrations of Hexanol 
at various temperatures. Right panel: Data collapse obtained using the scaling ansatz discussed in the text. The inset shows the 
growth of the length scale as a function of temperature in glass-forming Butanol.}
\label{exptButanol}
\end{center}
\end{figure*} 
In Fig.~\ref{exptGlycerol} (a), the variation of dielectric storage modulus with time is shown for $5\%$ concentration of sorbitol in the temperature range $203K$ to $248K$. Note that the unit of time is second, and the range of $\epsilon^{\prime}$ is re-scaled using the Havriliak-Negami fitting function \cite{HAVRILIAK_NEGAMI}. With increasing solute concentration, the relaxation becomes slower, which
proves that presence of larger Sorbitol molecules makes the dynamics slower. So these large particles can be expected to act as `soft pinning' sites for the glycerol molecules. In Fig. \ref{exptGlycerol}(b), the pinning susceptibility, calculated using Eqn ~\ref{pinSusExpt}, is shown. The maxima of $\chi_p(T,t)$ increases with decreasing temperature, which manifests the growth of amorphous order. This is obtained for $5\%$ sorbitol concentration in glycerol. A similar analysis for using smaller concentrations ($c<5\%$) also gives pinning susceptibility, but the data is noisier as the effect of sorbitol
on the dynamics at this small concentration is somewhat weak. Thus, we chose $c>5\%$ to compute the pinning susceptibility for our current experimental
data; remarkably, we find that pinning susceptibility directly yields the static correlation volume. $\chi_p$ obtained using $c = 7.5\%, 10\%$ and
$15\%$ yields the same value of the static correlation. This, we think, is a remarkable result as measuring the static correlation volume in any molecular
glass-forming liquids will now be very simple and reliable without many experimental difficulties. Next, we concentrate our attention on the scaling
analysis of relaxation time to test the correctness of the results via $\chi_p$ in a self-consistent manner. As we obtained relaxation time directly from
the peak position of the loss modulus, $\epsilon^{\prime\prime}$ data, we can perform the scaling analysis directly on the raw data. The relaxation
time of glycerol for various concentrations of sorbitol is shown in Fig. \ref{exptGlycerol}(c). Using the scaling argument discussed before
~\cite{RajsekharSoftMatter} we perform the scaling analysis as shown in Fig. \ref{exptGlycerol}(d). The inset of the same figure shows the data before
doing the scaling along the x-axis to extract the desired length scale. The data collapse is shown in Fig. ~\ref{exptGlycerol}(d) observed to be very
good, suggesting the good signal to noise ratio in experimental measurements. In Fig. ~\ref{exptGlycerol}(f), the measured length-scale is then
compared with the existing values reported in Ref.~\cite{BiroliScienceGlucerol-2} in which the fifth-order non-linear dielectric susceptibility
($\chi_{_5}(T,t)$) of Glycerol was measured and the length-scale is given as $\xi_s(T) \sim \left(\frac{\chi_{5}}{T^2}\right)^{1/6}$.
Again, the agreement is excellent to clear away any doubt about the reliability of this innovative method.

Once again, to reassure the accuracy of our experimental results, we analyzed some of the previously published data.
In Ref.~\cite{Duvvuri2004_GlycerolandSorbitol, Shachi_exptDataPRL}, the $\alpha$ relaxation time of Glycerol-Sorbitol mixture for various
concentrations are reported. Note that data for a small concentration of sorbitol are not presented in these studies. So, we have taken all the
reported data for all the concentrations (See SM for further details on the data extraction and consistency check of our data with the reported
ones) and combined our data for smaller concentrations to perform the scaling analysis over the entire range of data available with us.
In Fig.\ref{exptGlycerol}(e), we show the scaling analysis of the entire data set. Remarkably the data collapse
is found to be very good for all the data of various concentrations. The success of this grand scaling collapse suggests the robustness
of the scaling theory proposed in this work. In Fig.\ref{exptGlycerol}(f) we compare the length scale obtained from all the scaling analyses
along with the static length that can be obtained from peak height of $\chi_p$ as $\xi_s \sim \chi_p^{1/3}$ at various concentrations of sorbitol.
The comparison of the static length-scales measured using all these various ways and the reported value from Ref.~\cite{BiroliScienceGlucerol-2} are in good agreement with each other. Thus, our proposal of ``soft pinning'' to extract the
static length scale in glass-forming molecular liquids is robust beyond doubt, and we hope that a careful analysis of experimental data can
lead to an understanding of the growth of static length scales in various other molecular glass-forming liquids.

A similar analysis for Butanol using Hexanol as ``soft pinning'' molecules is presented in
Fig.\ref{exptButanol}. In the left panel of the figure, we have shown the extracted data of relaxation time of Butanol as a function
of the concentration of the Hexanol as soft pinning impurities for various temperatures (See SM for details of data extraction and other
analysis). In the right panel of Fig.\ref{exptButanol}, we show the data collapse obtained using the scaling ansatz Eq.\ref{pinningScaling}.
The underlying growth of the length scale is shown in the inset of the same figure. The data collapse in this Butanol-Hexanol mixture
is again found to be very good, suggesting that the extracted length scale will have smaller errors. Verification of these results using
other scaling methods is not possible at this moment as we do not have the $\epsilon^{\prime}$ and $\epsilon^{\prime\prime}$
data available with us for this Butanol-Hexanol binary mixture with varying compositions. The fact that scaling analysis works so
well even for this binary mixture of glass-formers supports this method's universality.

\section{Discussions and Conclusions}
We performed extensive computer simulations of model glass-forming binary mixtures liquids in two and three dimensions with
random pinning. We estimated the extent of correlation from the local pinning sites at each studied temperature to obtain the underlying
static correlation length. The obtained correlation length is found to be in close agreement with the static length-scale obtained using
other conventional methods like the Point-To-Set method. This establishes the usefulness of pinning sites in extracting the local correlations
in the medium. We then employed the idea of ``soft pinning'' sites using impurity particles with large diameters to see if they
can act as pinning sites up to the structural relaxation time of the host liquid medium. As shown previously in colloidal experiments
\cite{Ganapathi2018}, we also found that larger particles can be excellent pinning sites, and correlation around them can be used
to obtain the underlying correlation length. With this knowledge, we then developed a scaling theory to understand the dependence of relaxation time of the host liquid medium on ``soft pinning'' particles and measure the underlying static correlation
length of the system. This scaling analysis rationalizes all the data obtained from our simulations in a unified manner. We then performed
dielectric susceptibility experiments of glycerol in the supercooled temperature regime with varying concentrations of sorbitol as
soft pinning particles. We can also rationalize all our experimental data and the existing reported data from previous studies using the
same scaling theory. The correlation length extracted using this scaling analysis for glycerol is then compared with the results obtained
via measurement of fifth-order non-linear dielectric susceptibility from Ref.\cite{BiroliScienceGlucerol-2}. Excellent agreement between
these two length scales establishes the robustness of this simple yet elegant method. It opens up future possibilities to extract static
correlation length in various other glass-forming liquids to develop a unified understanding of the physics of glass transition.

\section{Method Section\label{Models}}

\subsection{Simulation Details}
We have performed extensive computer simulations of the following four well-studied model glass-forming liquids. 
 
\noindent{\bf (i) 3dKA} --  The first model that we studied is the well-known Kob-Andersen model \cite{KA1995}. It is a $80:20$ 
binary mixture of two type of  particles with interacting potential 
\begin{equation}
\label{LJ}
  V_{\alpha\beta}(r)=4\epsilon_{\alpha\beta}\left[\left(\frac{\sigma_{\alpha\beta}}{r}\right)^{12}-
    \left(\frac{\sigma_{\alpha\beta}}{r}\right)^{6}\right].
\end{equation}
The parameters of this model with $\alpha,\beta \in \{A,B\}$, are $\epsilon_{AA}=1.0$, $\epsilon_{AB}=1.5$,
$\epsilon_{BB}=0.5$, $\sigma_{AA}=1.0$, $\sigma_{AB}=0.80$,
$\sigma_{BB}=0.88$ and number density $\rho = 1.20$. The potential is cut off  at a distance $r_{cut} = 2.50\sigma_{\alpha\beta}$. We use a 
quadratic polynomial in such a way that the potential and its first two 
derivatives are continuous at the cut off. We study the system in a  temperature range
$T\in \{0.45,2.0\}$. The system size we chose is $N = 108000$.
Simulations with `soft pinning' particles are basically the ternary version of the same model, but $\alpha,\beta \in {A,B,C}$ with 
$\epsilon_{\alpha\beta} = \epsilon_{BC} = \epsilon_{CC} = 1.0$ and $\sigma_{AC} = 1.10$, $\sigma_{BC} = 1.04$ and 
$\sigma_{CC} = 1.20$ and all other parameters are same as above. 

\vskip +0.1in
\noindent{\bf (ii) 2dmKA} -- This two dimensional model is similar to the {\bf 3dKA} model but 
the particle number 
ratio is $65:35$. Such a number ratio is chosen to avoid any crystallization in the studied temperature ranges $T\in \{0.45,2.0\}$. The system 
size tin this case  is $N = 10000$. 

\vskip +0.1in
\noindent{\bf (iii) 3dR10} -- This model is a $50:50$ binary mixture \cite{KARMAKAR20121001} in three 
dimensions where the  particles interacts via 
\begin{equation}
\label{R10}
  V_{\alpha\beta}(r) = \epsilon_{\alpha\beta}\left(\frac{\sigma_{\alpha\beta}}{r}\right)^{10}.
\end{equation}
The parameters for this model are, $\epsilon_{\alpha\beta} = 1.0$, $\sigma_{AA} = 1.0$,
$\sigma_{AB} = 1.22$ and $\sigma_{BB} = 1.40$. Here the potential is cut off 
at a distance $r_{cut} = 1.38\sigma_{\alpha\beta}$ using a  similar quadratic 
polynomial so that the potential and its first two derivatives are 
continuous at the cutoff radius. The particles number density here is $\rho = 0.81$. 
The range of temperatures we study here is $T \in \{0.52,2.0\}$.  The system size for this model is
$N = 1000-108000$.
For soft pinning analysis we studied the ternary version of this model and use $\epsilon_{\alpha\beta} = 1.0$ 
and $\sigma_{CC} = 1.60$, $\sigma_{AC} = 1.30$ and $\sigma_{BC} = 1.50$ with all other parameters being 
kept same as above.
\vskip +0.1in

\vskip +0.1in
\noindent{\bf (iv) 2dR10} -- This model is similar to  {\bf 3dR10} model in two dimensions 
with number density $\rho = 0.85$. The temperature range for this model is 
$T \in \{0.520,2.000\}$ and the system size  is 
$N = 128-10000$. 

We perform molecular dynamics simulations in a constant number of particles ($N$), volume ($V$) and temperature ($T$)(NVT) 
ensemble for the binary models. The integration was done using a modified leap-frog algorithm. To keep the temperature constant during the simulations at each studied temperatures, we use a Berendsen thermostat. Note that any other thermostat does not change the results qualitatively. Before storing the data, we equilibrate the systems well enough -- at least run them for 
$100\tau_\alpha$. For the ternary model, we perform NPT molecular dynamics simulations using \textit{LAMMPS} software~\cite{LAMMPS}. 
The pressure at each temperature is chosen to be the pressure of the corresponding binary mixture at that temperatures. 
For each studied temperatures, we have performed $32$ independent realizations. For random pinning (RP), we first equilibrate the system and 
then chose a fraction $c = \tfrac{N_p}{N}$ ($N_p$ is number of pinned particles ) of particles and fix them at their  respective equilibrium positions. 
The rest of the particles are then evolved in time using normal Newtonian dynamics. For all the model systems, we use reduced units for the macroscopic variables. Lengths are measured in units of $\sigma_{AA}$, energy in units of $\epsilon_{AA}$ and time in units of 
$\sqrt{\tfrac{m\sigma_{AA}^2}{\epsilon_{AA}}}$ with $m$ being the mass of the particles which we have chosen to be $1$.

\subsection{Correlation Functions}
\noindent{\bf Overlap Correlation Function:}
To characterize the dynamics, we calculated the two-point density-density correlation function or the overlap function defined as below
\begin{equation}
Q(T,c,t) = \frac{1}{N-Nc}\left[\left\langle\sum_{i=1}^{N-Nc}w(|\vec{r}_i(t) -\vec{r}_i(0)|)\right\rangle\right],
\end{equation}
where the window function $w(x) = 1.0$ if $x \leq 0.30$ and $0$ otherwise and $N$ is the total number of particles and $Nc$ is 
number of pinned particles. $\left\langle....\right\rangle$ corresponds to the thermal averaging and $\left[...\right]$ corresponds to 
the averaging over different realizations of random pinning configurations. Any possible de-correlation of particles due to the rattling 
motion inside the cage formed by neighboring particles is removed by the window function. The relaxation time $\tau_\alpha$ is defined 
as $Q(t = \tau_\alpha) = \tfrac{1}{e}$.  For the pinned system, however we calculate the overlap function for only the mobile particles.
\vskip +0.1in
\noindent{\bf Four-point Correlation Function:}
The four-point susceptibility which estimates the degree of dynamic heterogeneity is defined as the 
fluctuations~\cite{Karmakar2014,KDS2009} in the overlap function and is given by,
\begin{equation}
\chi_4(t) = N \left[ \left\langle Q^2(t)\right\rangle -\left\langle Q(t)\right\rangle^2 \right].
\end{equation}
The four-point susceptibility defined above generally estimates the overall extent of dynamic heterogeneity of the system rounded by 
the finite system size.

\subsection{Experimental Details}
\begin{figure}[!htp]
\begin{center}
\includegraphics[scale =0.38]{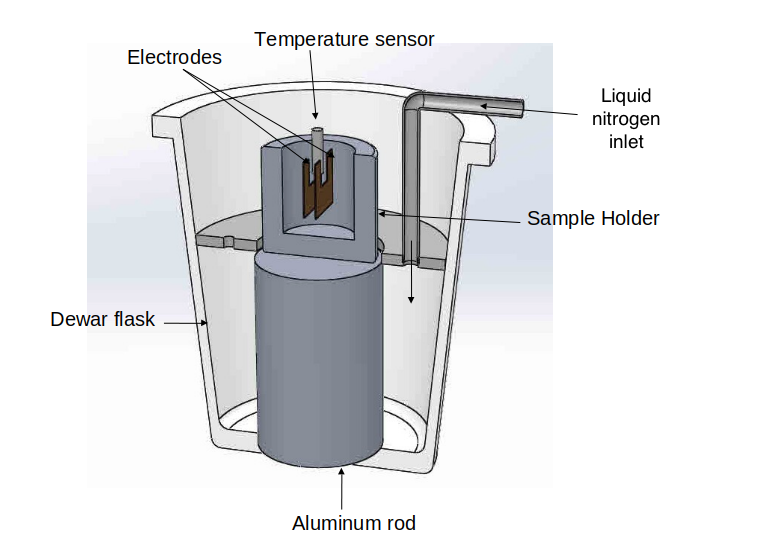}
\caption{Schematic presentation of the home-built experimental setup containing a dewar flask which contains liquid nitrogen 
during experimental runs. An aluminum rod connected to the sample chamber is used for sample chamber cooling. 
A two-electrode plate connected to a potentiostat (Biologic SP300) is used for dielectric spectroscopy at different sample 
temperatures. }
\label{setup}
\end{center}
\end{figure}

The schematic of the experimental setup used in this work is shown in Fig.~\ref{setup}. To calculate the dielectric modulus, we used a 
parallel plate capacitor kept inside an Aluminum cup which we call a sample holder. The electrodes of the capacitor are made of stainless steel to avoid any kind of chemical reaction with the samples. Teflon plates are used to avoid contact between electrodes and the sample holder. The separation between two electrodes is $1 mm$, and the area of the electrodes is $36cm^{2}$. The sample holder was connected to a solid massive Aluminum rod of very large heat capacity. This whole structure was kept inside a Dewar flask to keep it thermally isolated from the environment.

\vskip +0.1in
\noindent{\bf A. Dielectric Spectroscopy Experiments: }
We performed dielectric spectroscopy of ultra-pure Glycerol and Sorbitol purchased from Sigma and used as such.
Glycerol is a strong glass former and exhibits glass transition at $193K$. We used Sorbitol as solute particles. Sorbitol molecule has a liner size that is roughly double of a Glycerol molecule. Thus it should 
have a diffusion coefficient significantly smaller than Glycerol and can act as soft pinning  particles in supercooled 
Glycerol medium. The dielectric spectroscopy of Glycerol and Sorbitol mixture is studied in 
\cite{Duvvuri2004_GlycerolandSorbitol}. We prepare the sample by mixing the appropriate amount of Glycerol and 
Sorbitol according to their atomic mass at room temperature. While making the sample, special care has been taken so that samples do not get any contamination. The obtained data is cross verified with the previously reported
data in Ref.~\cite{Duvvuri2004_GlycerolandSorbitol} (see SM for details).

\vskip +0.1in
\noindent{\bf B. Data collection: }
The sample was poured inside the sample holder, which contains the parallel plate capacitor. After that, the heavy aluminum rod was cooled using liquid nitrogen. As the aluminum cup and the sample inside it were connected, it cooled the samples automatically. The temperature of the sample was monitored using an electronic temperature detector. 
Once the temperature reached the desired temperature, we stopped cooling. After that, the temperature starts to increase due to the heat transfer to the environment. As the whole system was suitably thermally insulated, the rate of this increment was low, which allowed us to have enough time to collect the data. Using this method at each temperature window, we were able to maintain 
the temperature for around 3-4 minutes to perform our measurements. 
We consider dielectric storage modulus ($\epsilon^\prime$) as the suitable quantity to measure the proposed pinning susceptibility 
~\cite{BiroliScienceGlucerol, BiroliScienceGlucerol-2}. We stress that as $\epsilon^\prime(t)$ and $q(t)$ are both two point correlation 
functions, one can thus compute the pinning susceptibility from $\epsilon^\prime(t)$ as well. The susceptibility is then defined as  
\begin{eqnarray}
\chi_p^{Expt}(T,t) &=& \left .\left(\frac{\partial \epsilon^{\prime}(T,c,t)}{\partial c}\right)\right|_{c=0} \nonumber\\ 
&=&  \left. \frac{\epsilon^{\prime}(c, T, t) - \epsilon^{\prime}(c=0, T, t)}{c}\right|_{c=0},
\label{pinSusExpt}
\end{eqnarray}
where $c$ is the concentration of Sorbitol. We measured $\epsilon^\prime$ of Glycerol-Sorbitol mixture for 
a range of temperature $T\in \{203K ... 248K\}$. We varied Sorbitol concentration ($c$) from $0$ to $15\%$. 
The range of $\epsilon^\prime$ is re-scaled within $0 - 1$ using a modified form of Havriliak-Negami fitting function for the 
dielectric relaxation\cite{HAVRILIAK_NEGAMI} (see SM for details).  Note that the fitting procedure is used only
to obtain the low and high frequency limit of the storage modulus for the normalization purpose. No fitting is involved 
any other part of the experimental data analysis.

\vskip +0.1in
\noindent{\bf Authors' contributions :}
SK conceived the research project. RD did all the theoretical and simulation works. BPB, APB and TNN performed the experiments. SK and RD
did all the data analyses. RD and SK wrote the paper in consultation with other co-authors.

\vskip +0.1in
\noindent{\bf Acknowledgments:}
We thank Narayanan Menon for his help regarding the experimental setup. TIFR ballon facility, Hyderabad, is acknowledged for providing 
the workshop facility. This project is funded by intramural funds at TIFR Hyderabad from the Department of Atomic Energy (DAE). 
Support from Swarna Jayanti Fellowship grants DST/SJF/PSA-01/2018-19 and SB/SFJ/2019-20/05 are also acknowledged. 

\bibliography{sciadvfile}
\bibliographystyle{ScienceAdvances}

\end{document}


\title{Soft-Pinning: Experimental Validation of Static Correlations in Supercooled Molecular Glass-forming Liquids
--supplementary materials. }
\author{Rajsekhar Das$^{1,2}$}
\author{Bhanu Prasad Bhowmik$^{1,3}$}
\author{Anand B Puthirath$^{1,4}$}
\author{Tharangattu N. Narayanan$^{1}$ }
\author{Smarajit Karmakar$^{1}$}
\email{smarajit@tifrh.res.in}

\affiliation{$^1$Centre for Interdisciplinary Sciences, Tata Institute of Fundamental Research, 
36/P, Gopanpally Village, Serilingampally Mandal,Ranga Reddy District, Hyderabad, 500107, India}
\affiliation{$^2$ Department of Chemistry, University of Texas at Austin, Austin, Texas 78712, USA}
\affiliation{$^3$ Weizmann Institute of Science, Rehovot, Israel}
\affiliation{$^4$ Department of Materials Science and Nano Engineering,
Rice University, 6100 Main Street, Houston, TX, 77005, USA}

\maketitle

\section{Point-To-Set Method}
In the main text, we have compared the length scales obtained using the new method with the one obtained using the well-known Point-To-Set(PTS)~\cite{Biroli2008} method, described below. This method was originally introduced by Bouchaud and Biroli~\cite{JPBBG}. The idea is first to equilibrate the system at the desired temperature. Then one needs to define a region of radius $R$ outside which all the particles are pinned at their respective equilibrium positions. The particles inside the spherical region are then allowed to equilibrate via natural dynamics. For small values of $R$, the particles inside the spherical region will remain correlated due to the frozen boundary effect created by the outside particles. They will not be able to decorrelate even at large enough time. The decorrelation will only happen when the radius $R$ of the cavity is allowed to be sufficiently large compared to the inherent length scale $\xi_{PTS}$ of the system. The static length scale is then identified as the value of $R$ where this crossover occurs. To systematically characterize it, one needs to calculate a static overlap correlation function defined as
\begin{equation}
q_c(R) = \lim_{t\to\infty} \frac{1}{ml^3\rho} \sum_{i=1}^N\left\langle n_i(0)\right\rangle\left\langle n_i(t)\right\rangle,
\end{equation}
where $m$ is the number of small boxes of linear size $l$ in the central region of the cavity, $\rho$ is the number density of liquid and $\left\langle...\right\rangle$ denotes the averaging. $n_i(t) = 1$ if a box is occupied by a particle at time $t$ and $n_i(t) = 0$ 
if it is empty. The $l$ is chosen in such a way that one box can accommodate only one particle. The function is calculated in the 
limit $t\to \infty$. The function will decay from $1$ to the bulk value $q_{0} = \rho l^3$, the probability that a box will be occupied. 
For each value of $R$ this is then fitted to an exponential function $q_c(R)-q_{0} = A\exp\left[-\left(\frac{R-1.0}{\xi_{PTS}}\right)^{\eta}\right]$ 
to extract the length scale $\xi_{PTS}$ (for details see~\cite{PhysRevLett.111.165701}).  In the main text  some of the length scales 
compared was taken from Ref.~\cite{RD2016}.
\section{Configurational overlap}
\begin{figure}[!h]
\begin{center}
\includegraphics[width=0.97\textwidth]{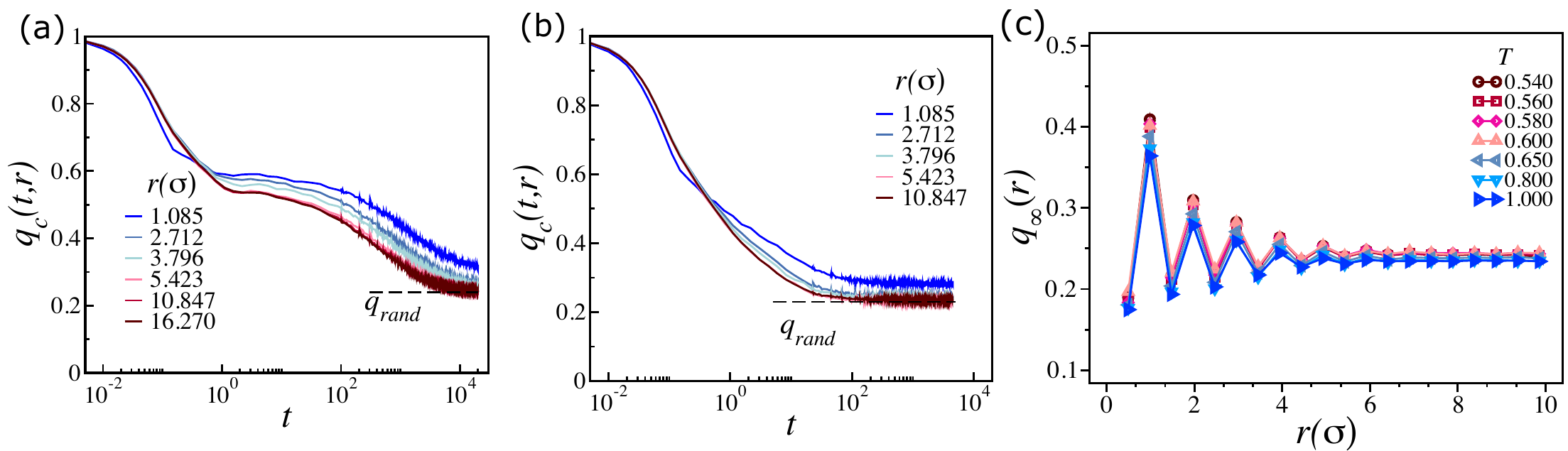}
\caption{ Panel (a) and (b) shows the variation of $q_c(t,r)$ as a function of $t$ for two different temperatures $T= 0.600$ and $T = 1.000$ respectivley for the \textbf{2dR10} model. Panel (c) shows $q_{\infty}$ as a function of $r$ from the pinning site.}
\label{qctr}
\end{center}
\end{figure}  
Configurational overlap $q_c(t,r)$ described in main text is shown at various values of $r$ for two different temperatures ($ T = 0.600$ and $T = 1.00$) for \textbf{2dR10} model in panel (a) and (b) of Fig.~\ref{qctr}. For high temperature the function decays very fast and reach the bulk value. Whereas for the low temperature ($T = 0.600$) we see the basic feature of glassy dynamics -- the two step relaxation. The $q_{\infty}(r)$ for the same model is shown in panel (c).

\section{Relaxation time as a function of radial distance ($r$) from the pinning site and Scaling analysis}
\begin{figure}[!h]
\begin{center}
\vskip - 0.1in
\includegraphics[width=0.85\textwidth]{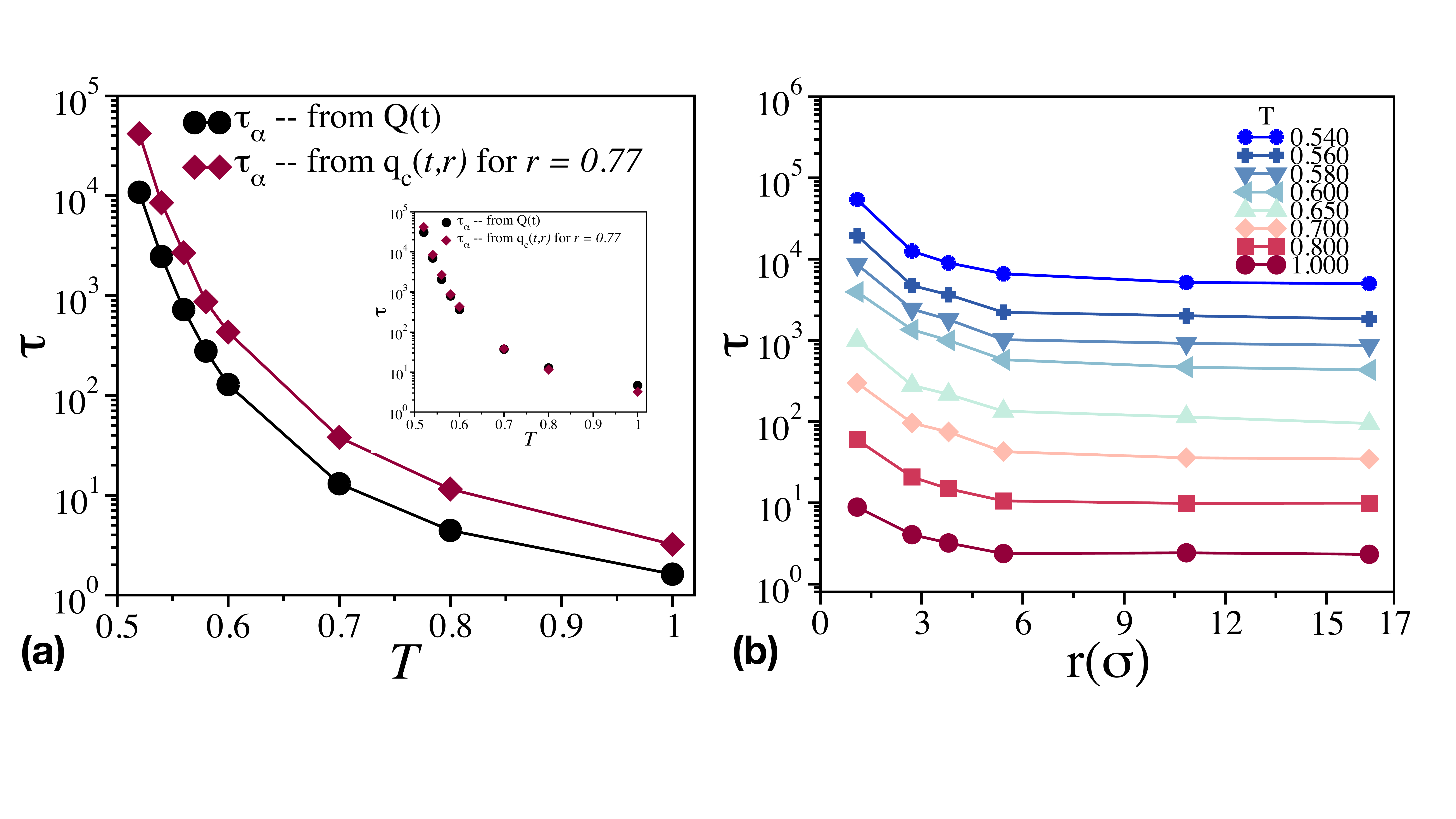}
\vskip - 0.5in
\caption{(a) Comparison of the temperature dependence of the relaxation time at a certain radial distance from a pinning site with the bulk relaxation time for the \textbf{3dR10} model. Inset shows the scaled version to show that they have almost the same temperature dependence. (b) Shows the dependence 
of relaxation time of particles situated at a radial distance $r$ from a pinning site for all the studied temperatures for the 2dR10 model.}
\label{timescales}
\end{center}
\end{figure}

In Fig.\ref{timescales}a, we have shown the temperature dependence of relaxation time at a distance $r=0.77\sigma$ from the pinning site
with the bulk relaxation time for \textbf{3dR10} model. It is indeed true that for all temperatures, the relaxation time near the pinning site is larger than the bulk value, but the temperature dependence is very similar to the bulk relaxation time, as shown in the inset of the same figure. In Fig.\ref{timescales}b,
we have demonstrated the radial dependence of average relaxation time ($\tau_\alpha(r, T)$) computed from a pinning site for various temperatures for the 2dR10 model. 
The radial dependence of $\tau_\alpha(r, T)$ clearly shows that with lowering the temperature, the effect of the pinning site is felt up to a longer distance, indicating the interplay of the underlying static length scale.

\begin{figure}[h!]
\begin{center}
\includegraphics[width=0.93\textwidth]{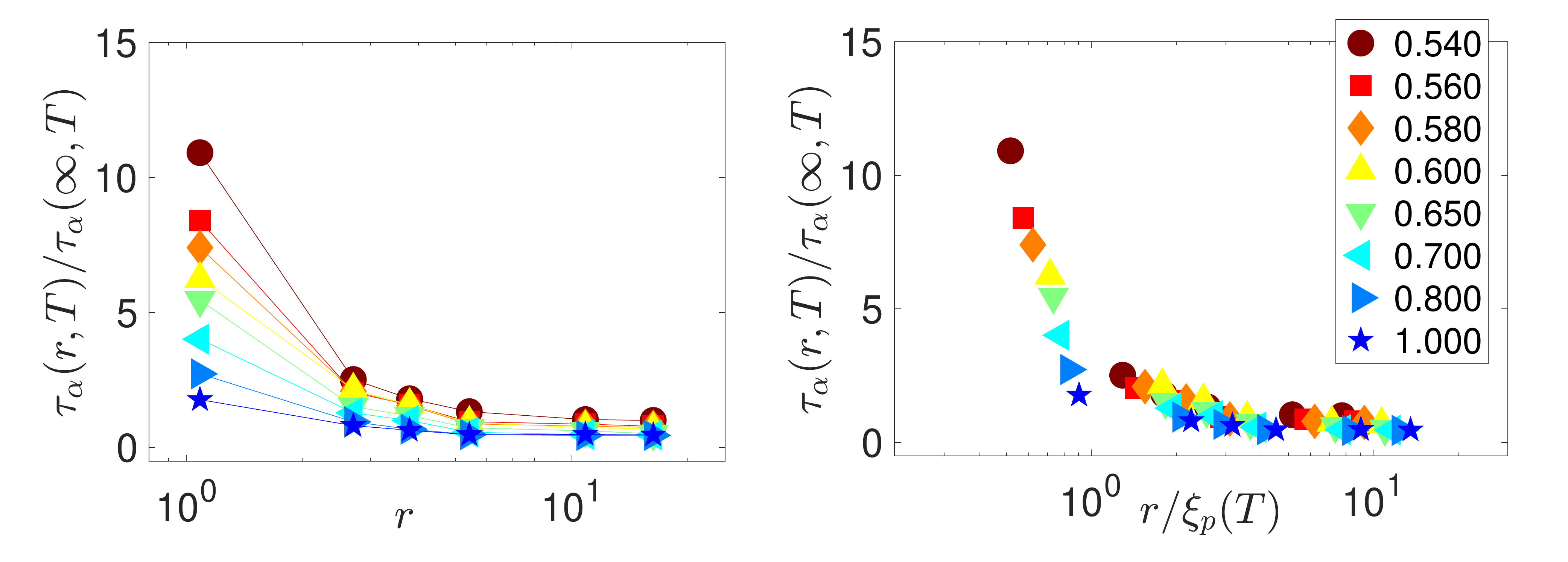}
\vskip - 0.2in
\caption{(a)$ \tfrac{\tau_\alpha(r,T)}{\tau_\alpha(\infty,T)}$ is plotted as a function of $r$ from the pinning site. Here $\tau_\alpha(\infty,T)$ is 
chosen to be same as the bulk value of $\tau_\alpha(T)$.  (b)  Data is collapsed using the same pinning length scale obtained from the decay of 
$|q -q_{rand}|$ (see Figure.1 in the main text). Note that in this data collapse, we do not have any adjustable parameter. Even then, the data collapse is found to be quite good. This suggests that the spatial dependence of relaxation time is indeed governed by the underlying static length scale of the system at that particular temperature.}
\label{scalingtauvsr}
\end{center}
\end{figure}

As discussed in the main article, if the relaxation time of particles that are $r$ distance away from a pinning site and the relaxation dynamics of
the system is controlled by an underlying static length scale, the one can expect that the spatial dependence of relaxation time from the pinning site 
will obey the  following scaling relation
\begin{equation}
\tau_\alpha^p(r,T) = \tau_\alpha(T)\mathcal{F}\left[ \frac{r}{\xi_s(T)}\right],
\label{Assumption}
\end{equation}
where $\xi_s$ is the underlying static length scale and $\mathcal{F}(x)$ is a scaling function which goes to $1$ as $x\to\infty$. 
This assumption is used in the main article to derived the final scaling relation between the relaxation time of the whole system
in the presence of pinning particles and the underlying static length scale as given by
\begin{equation}
\frac{\tau_\alpha(c,T)}{ \tau_\alpha(T) } \simeq 1 + \kappa c\xi_s^3 \quad \mbox{or} \quad \log{\left[ \frac{\tau_\alpha(c,T)}{ \tau_\alpha(T) } \right]} \simeq \kappa c\xi_s^3,
\label{scalingAnsatz}
\end{equation}
where $\kappa = \rho\int_{0}^{\infty}[\mathcal{F}(x) - 1]x^2dx$. Thus, the validity of the scaling relation scaling relation in 
Eq.\ref{scalingAnsatz} crucially depends on the validity of Eq.\ref{Assumption}. If the assumption made in Eqn.~\eqref{Assumption} 
is correct then one might be able to do a scaling collapse of the $r$ dependent $\tau_\alpha$ data shown in Fig.~\ref{timescales} panel (b)  
using the static length scale. To check this we do the scaling analysis in 2dR10 model with random pinning. It is important to note that in the 
presence of random pinning the appropriate static length scale is the pinning length scale, $\xi_p$ ~\cite{RD2016}. The data collapse using 
the $\xi_p$ in 2dR10 model is shown in the Fig.~\ref{scalingtauvsr}a and Fig.~\ref{scalingtauvsr}b.
The data collapse obtained in Fig.~\ref{scalingtauvsr}b is reasonably good without any adjustable parameters suggesting the validity of the scaling assumption, Eq.\ref{Assumption} and the robustness of the scaling relation Eq.\ref{scalingAnsatz}. One can see that collapse is bad for
$T=0.8$ and $1.0$, which are high temperatures, and measurement of spatial dependence of relaxation time at these high temperatures can be
noisy as correlation length will be very small.

\section{Extraction of existing experimental data from literature}
\label{Method}
\begin{figure}[h!]
\begin{center}
\includegraphics[width=0.85\textwidth]{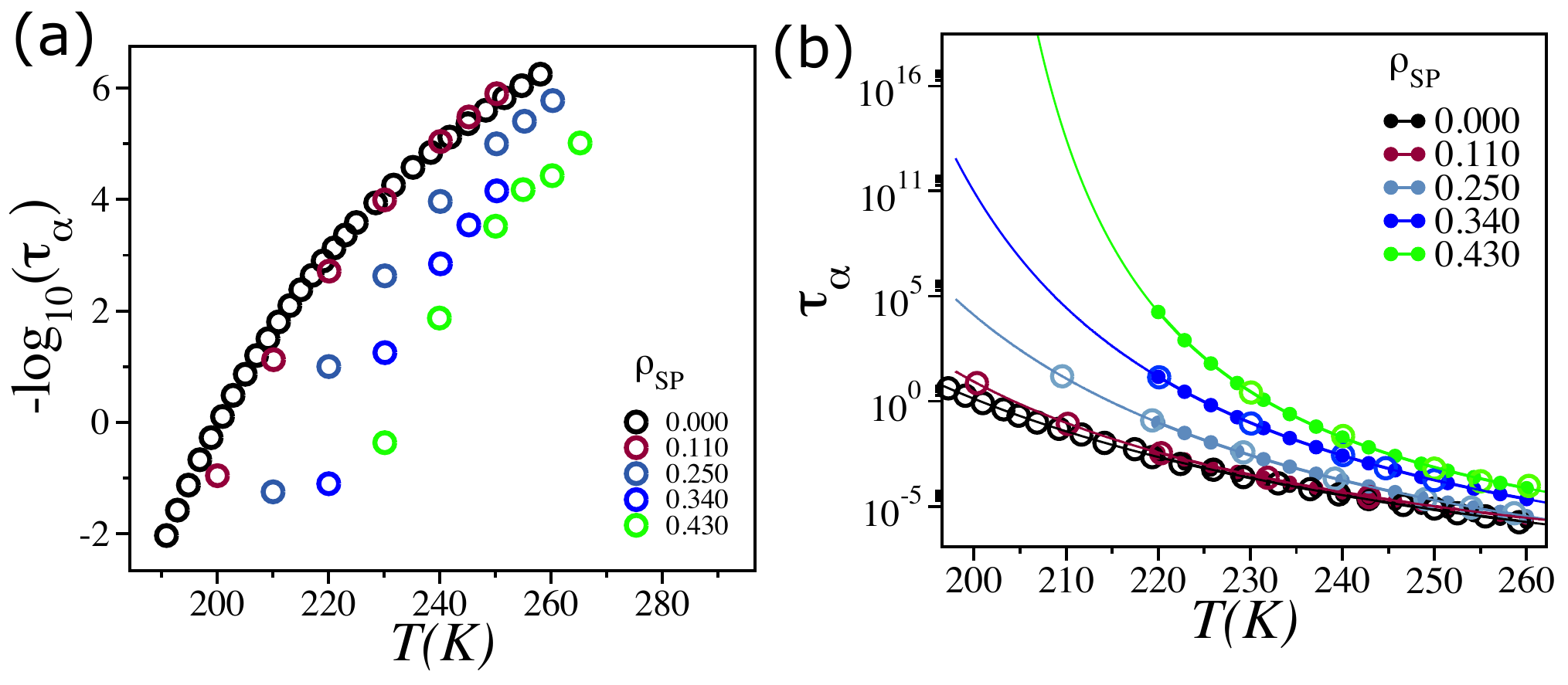}
\caption{(a)-- Raw data extracted from Figure.2\cite{Duvvuri2004} for Glycerol-Sorbitol mixture. (b) Shows  the method to extract $\tau_\alpha$ 
at desired temperatures using $VFT$ fitting of the raw data. }
\label{PROCEDURE1}
\end{center}
\end{figure}  
\vskip +0.2in
We use our scaling relation to extracting length scales for numerical and experimental systems in the main text.  We studied two different experimental systems --(i) Glycerol-Sorbitol mixture and (ii) Butnaol-Hexanol mixture. For Glycerol and Sorbitol mixture, we perform an experiment in our lab as well as use data extracted from previously published experimental data. We use WebPlotDigitizer software to extract raw data from the published paper. The Glycerol Sorbitol data was extracted from\cite{Duvvuri2004}. We first extracted the raw data from figure.2 of Ref.\cite{Duvvuri2004}. We then fitted the $T$ vs $\tau_\alpha$ data to the VFT equation given by
\begin{equation}
\tau_\alpha(T) = \tau_0\exp\left(\frac{AT}{T-T_{VFT}}\right).
\label{VFT}
\end{equation} 
The $VFT$ fitted data was used to extract time-scales ($\tau_\alpha$) at desired temperatures for all the studied fraction 
$\rho_{SP}$ Sorbitol in the Glycerol. We show the complete procedure in Fig.~\ref{PROCEDURE1}. Panel (a) shows the raw data. 
In panel (b), we show the $VFT$ fit of the data to \eqref{VFT}. The open circle represents the raw data, and the solid lines are the fit to the equation  
\eqref{VFT}. The small filled circle represents the extracted data at desired temperatures.
\begin{figure}[h!]
\begin{center}
\includegraphics[width=0.85\textwidth]{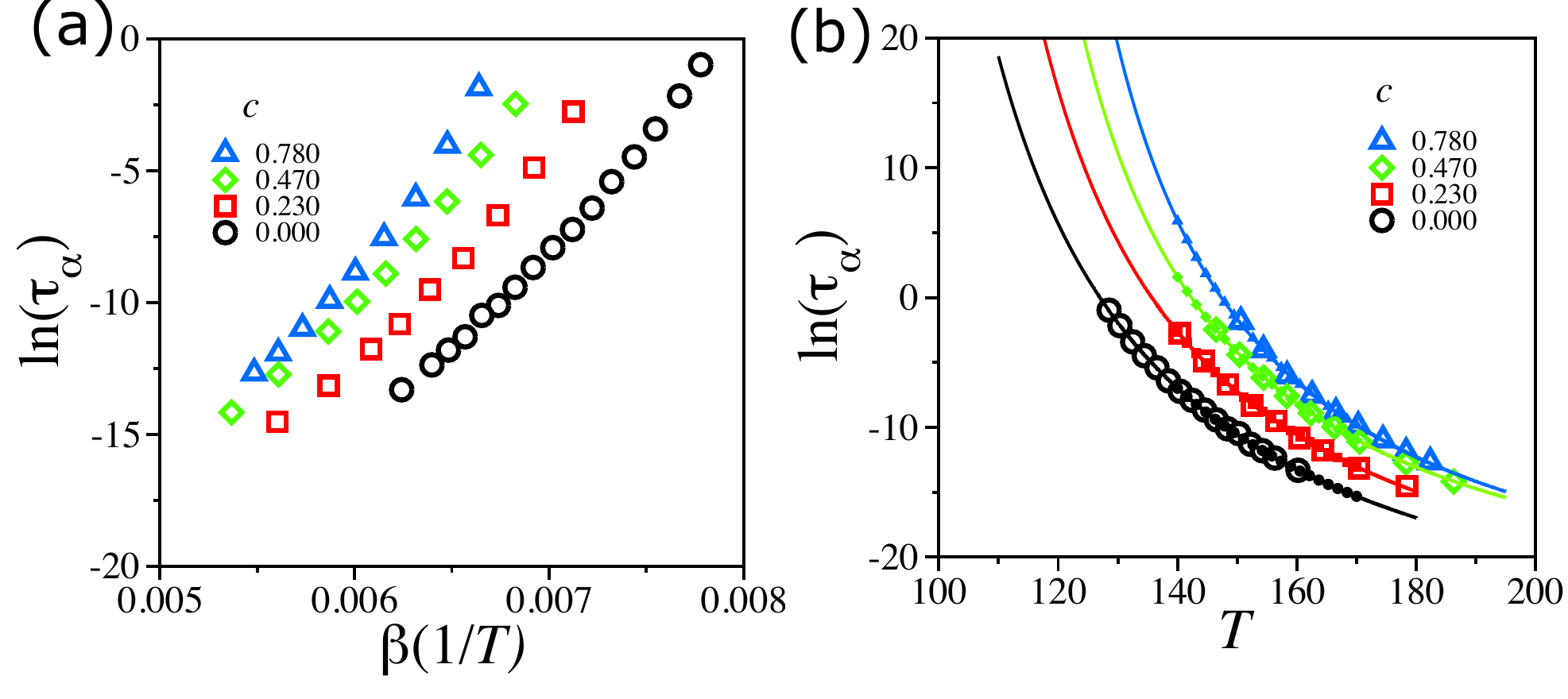}
\caption{(a)-- Raw data extracted from Figure.3 (a)\cite{Shachi_exptDataPRL} for Butanol-Hexanol mixture. (b) Shows methods to extract $\tau_\alpha$ at desired temperatures using $VFT$ fitting of the raw data. }
\label{PROCEDURE1-2}
\end{center}
\end{figure}  
We follow the same method for extracting the Butanol and Hexanol mixture shown in the main text from the Ref.\cite{Shachi_exptDataPRL}. 
The complete extraction methods are shown in Fig.~\ref{PROCEDURE1-2}. Note that for this mixture, we do not have data for small concentrations of Hexanol in Butanol. Nevertheless, we used this data to see if the dependence of the relaxation time of the concentrations of Hexanol in Butanol can be rationalized using the same argument that is employed to understand the data of Glycerol-Sorbitol mixture. The results suggest
that the scaling arguments hold even for this data set, leading to the extraction of the underlying static length scale. We could not independently
check the validity of this obtained length scale, and future studies, particularly at a smaller concentration of Hexanol in Butanol, will be needed.

\section{Dielectric Loss and Storage Modulus of Glycerol+Sorbitol Mixture}
\begin{figure}[h!]
\begin{center}
\includegraphics[width=0.9\textwidth]{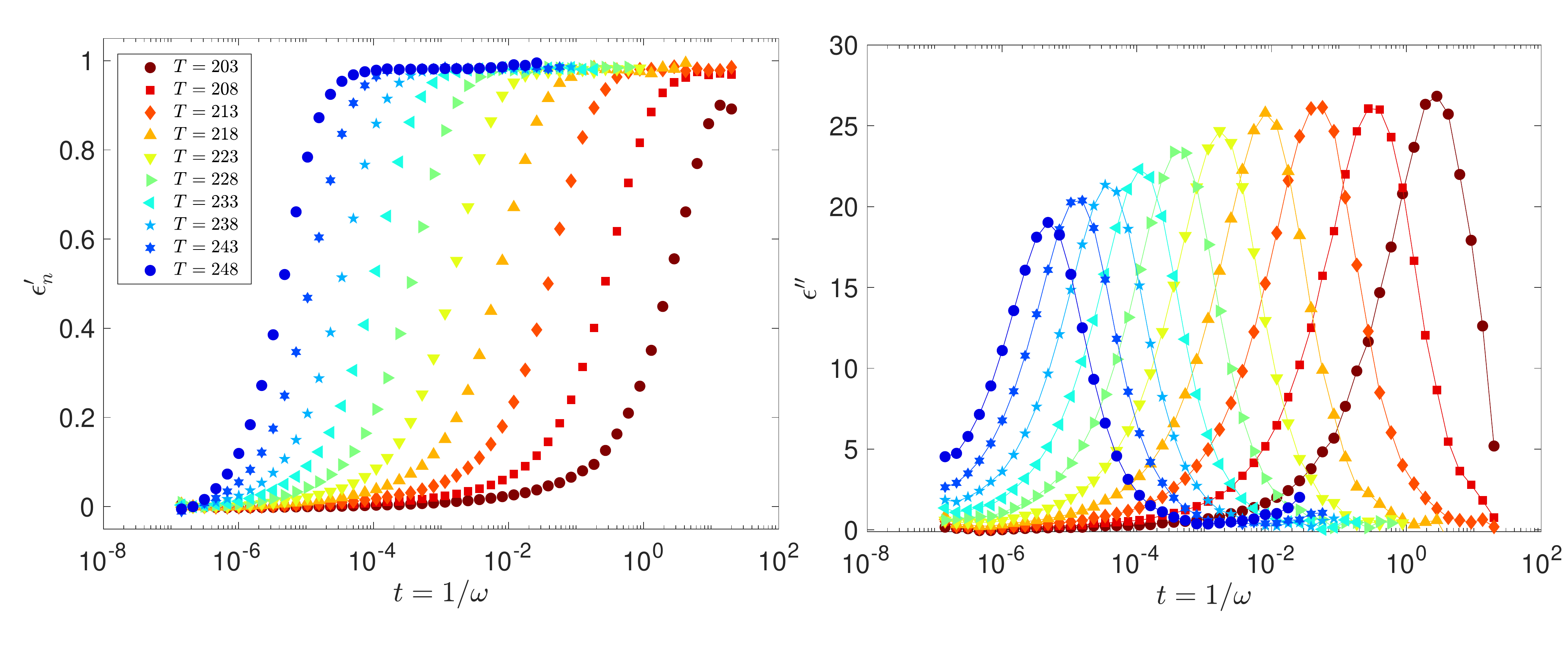}
\includegraphics[width=0.9\textwidth]{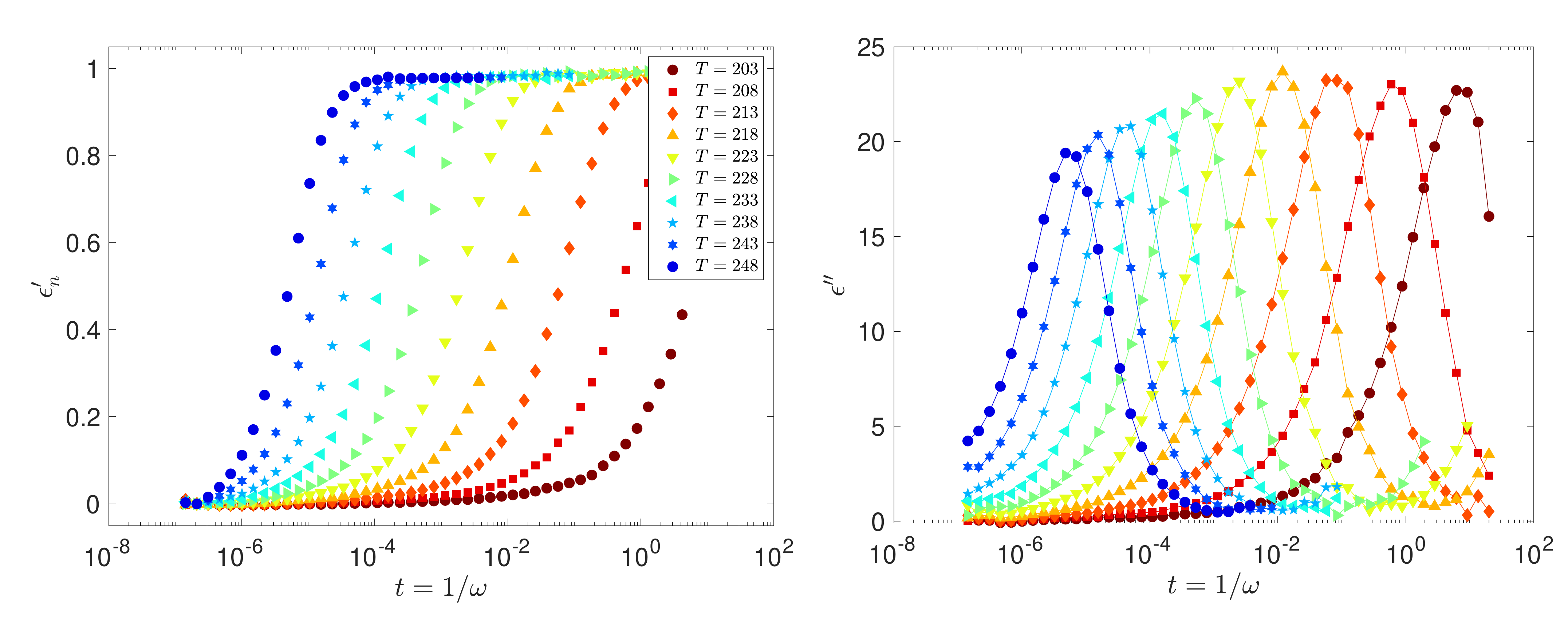}
\includegraphics[width=0.9\textwidth]{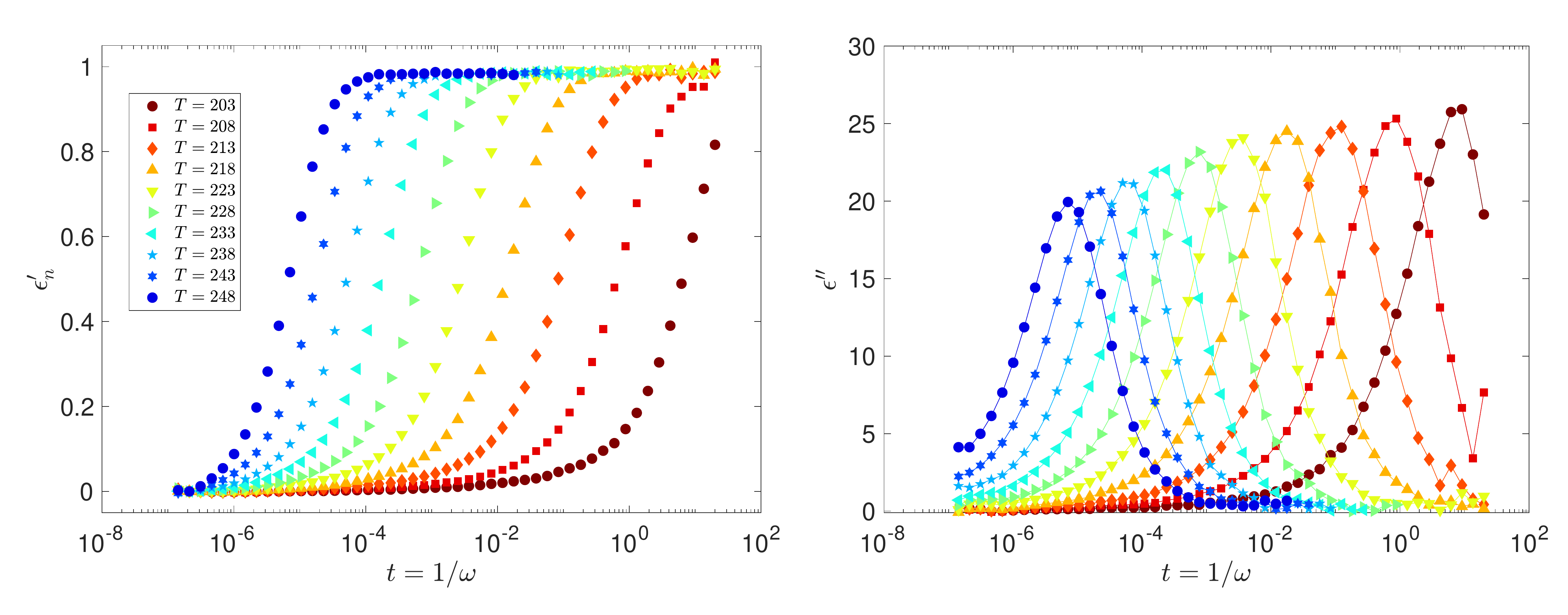}
\caption{Top panels: Normalized storage modulus ($\epsilon^{\prime}$) and loss modulus ($\epsilon^{\prime\prime}$) for pure Glycerol sample.
Middle panels: Similar data for Glycerol + $2\%$ Sorbitol mixture. Bottom panels: For Glycerol + $5\%$ Sorbitol mixture }
\label{dielectricEpsilon1}
\end{center}
\end{figure}

\begin{figure}[!htpb]
\begin{center}
\includegraphics[width=0.95\textwidth]{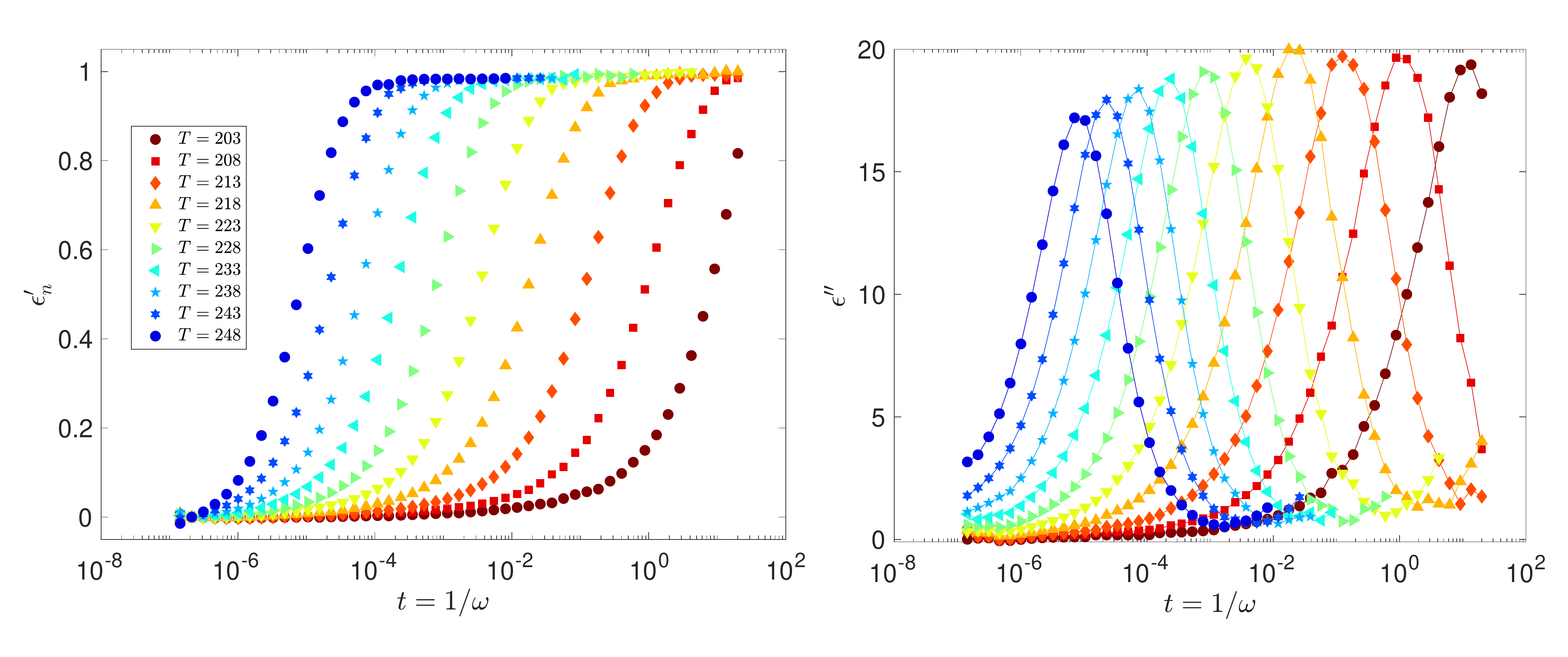}
\includegraphics[width=0.95\textwidth]{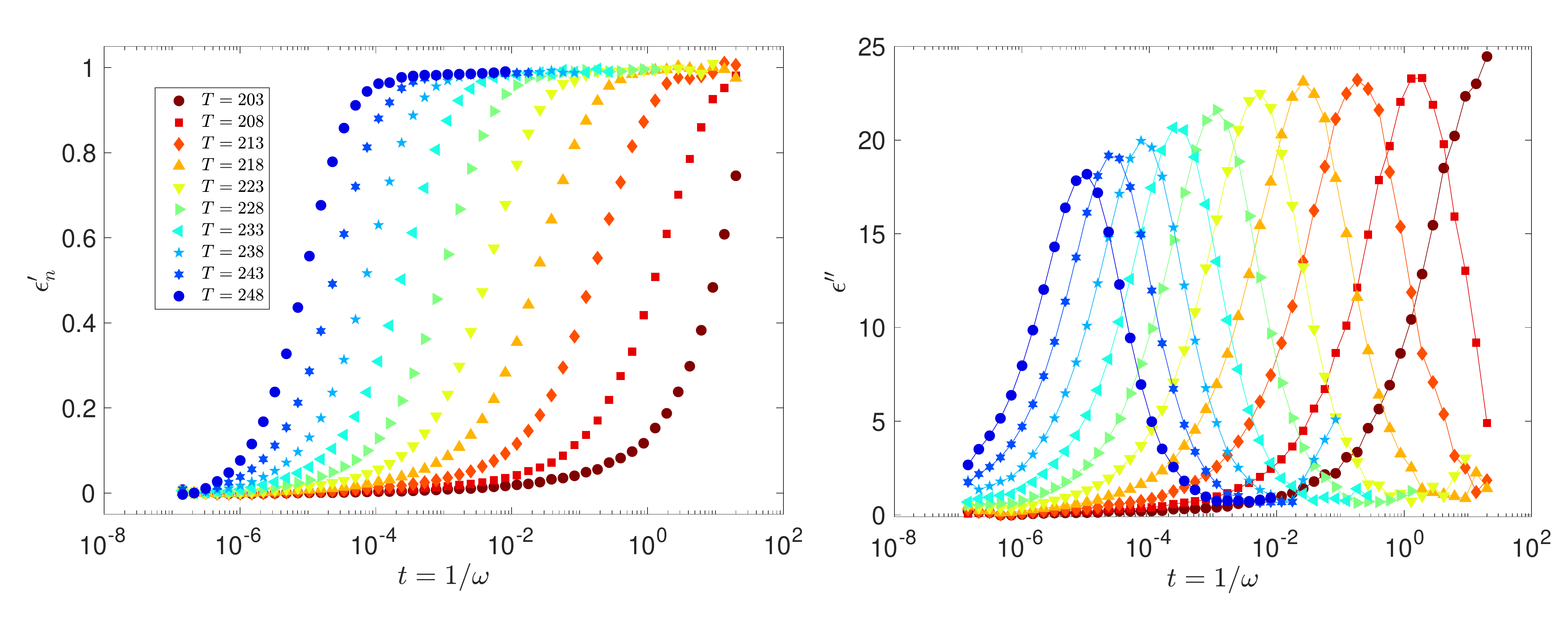}
\includegraphics[width=0.95\textwidth]{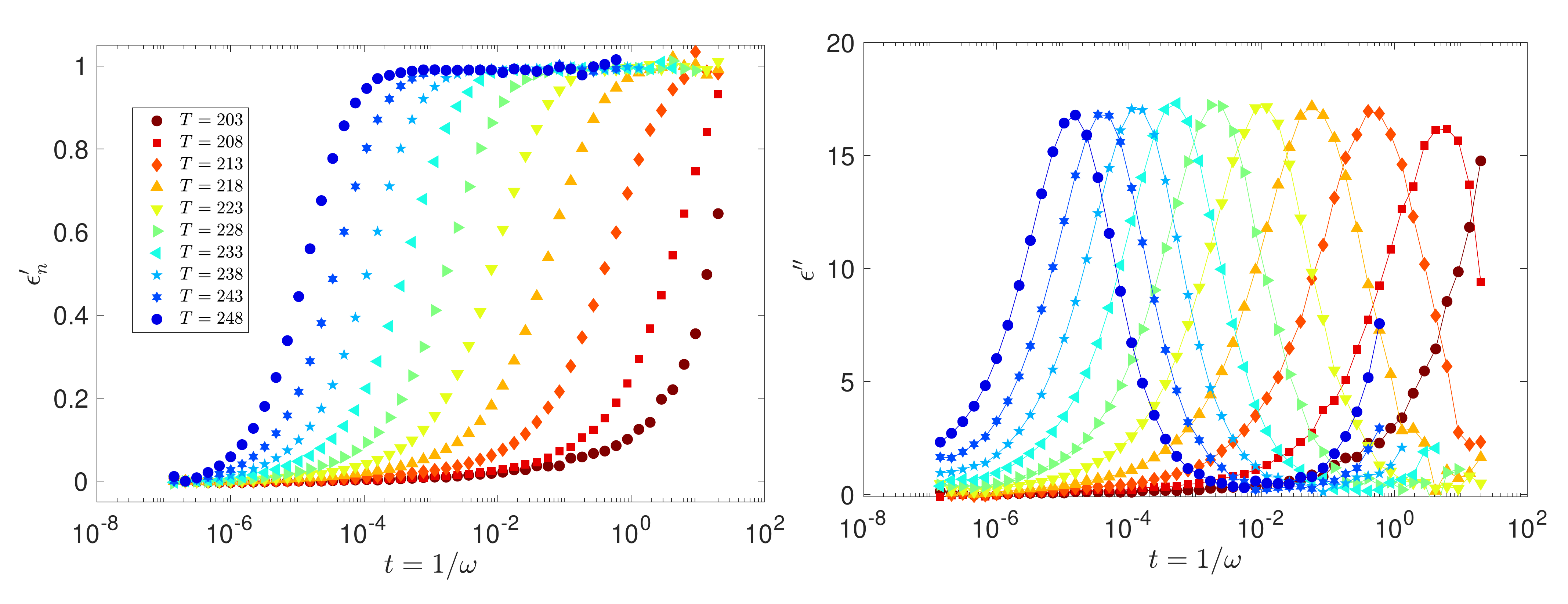}
\caption{Top panels: Normalized storage modulus ($\epsilon^{\prime}$) and loss modulus ($\epsilon^{\prime\prime}$) for 
Glycerol + $7.5\%$ Sorbitol mixture sample.
Middle panels: Similar data for Glycerol + $10\%$ Sorbitol mixture. Bottom panels: For Glycerol + $15\%$ Sorbitol mixture}
\label{dielectricEpsilon2}
\end{center}
\end{figure}
In this section, we have given all the data of storage and loss modulus ($\epsilon^{\prime}$ and $\epsilon^{\prime\prime}$) for the pure glycerol and the glycerol-sorbitol mixture with varying concentration of sorbitol. Note that the data of storage modulus ($\epsilon^{\prime}$) is normalized 
following modified version of Havriliak-Negami fitting formula \cite{HAVRILIAK_NEGAMI} as given below 
\begin{equation}
\epsilon^{\prime}(\omega) = \epsilon_\infty+\frac{\epsilon_s-\epsilon_\infty}{1 + (\omega\tau)^\alpha},
\end{equation}
where $\epsilon_\infty$ and $\epsilon_s$ are the high and low frequency limits of $\epsilon^{\prime}$. $\tau$ is the typical relaxation time and
$\alpha$ is a fitting exponent. This form is simplified, and we highlight that it fits the data quite well. Note that this fitting is used only to get 
$\epsilon_\infty$ and $\epsilon_s$ for normalizing the data to the range $[0,1]$. This normalization is needed to compute the pinning susceptibility
$\chi_p \equiv \frac{\partial \epsilon^{\prime}}{\partial c}$, where $c$ is the concentration of the co-solvent (Sorbitol in Glycerol-Sorbitol mixture and Hexanol in Butanol-Hexanol mixture). Any other data analysis does not involve the results obtained from this fitting procedure. 
In the left panels of Figs.\ref{dielectricEpsilon1} and \ref{dielectricEpsilon2}, we have plotted 
$\epsilon^{\prime}_n = \frac{\epsilon - \epsilon_\infty}{\epsilon_s - \epsilon_\infty}$ for various concentrations of Sorbitol in supercooled 
Glycerol.

We also highlight that one can obtain the relaxation time of the liquid mixture from this fitting procedure, but we have rather used the
peak value of $\epsilon^{\prime\prime}$ to determine the structural relaxation time, $\tau_\alpha$ just to avoid any error that might come due to fitting procedure. Also, note that to get a very precise value of the peak position in $\epsilon^{\prime\prime}$, we have
fitted to the peak of the loss modulus using a quadratic function and then determined the maximum position and the corresponding time (frequency) where the peak appears. This also removes any human intervention while estimating the relaxation time, $\tau_\alpha$
of the Glycerol-Sorbitol mixture from the peak position of the loss modulus. In the right panels of Figs.\ref{dielectricEpsilon1} and 
\ref{dielectricEpsilon2}, we have plotted $\epsilon^{\prime\prime}$ for all the studied concentrations of Sorbitol in Glycerol.
\clearpage

\section{Pinning Susceptibility for various concentrations of Sorbitol}
\begin{figure}[h!]
\begin{center}
\includegraphics[width=0.91\textwidth]{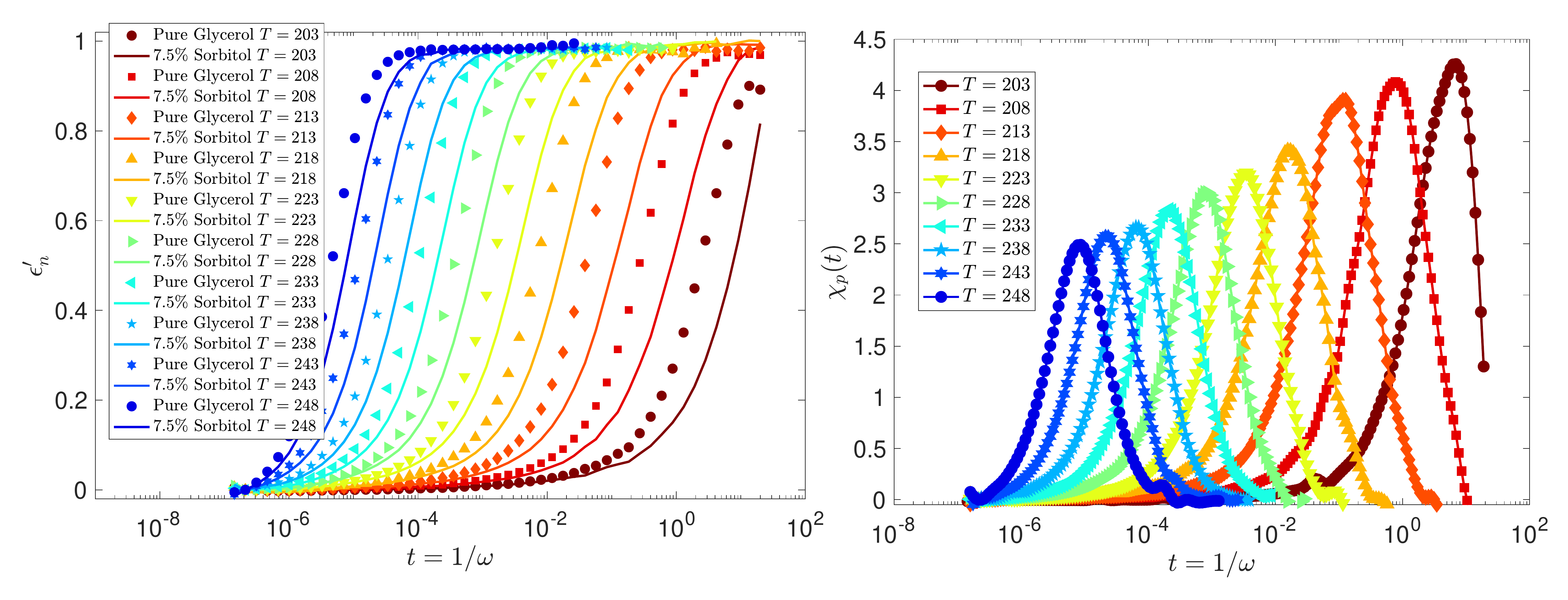}
\includegraphics[width=0.91\textwidth]{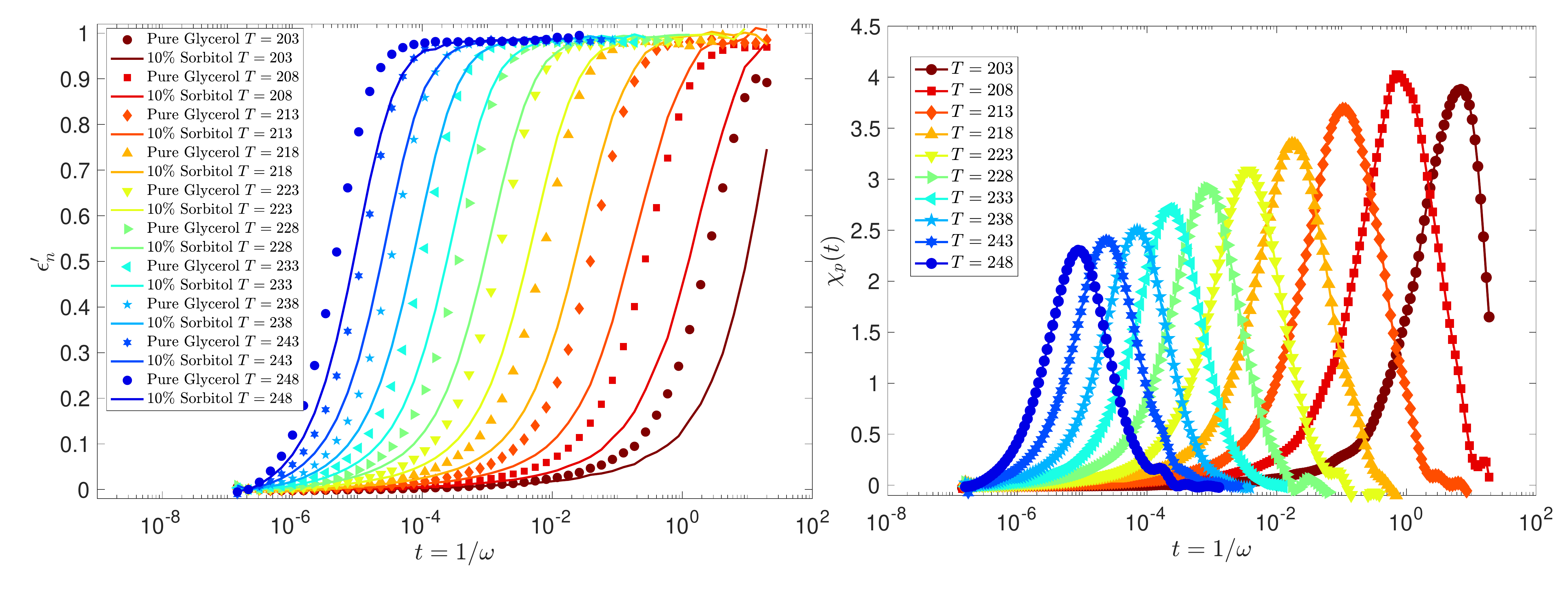}
\includegraphics[width=0.91\textwidth]{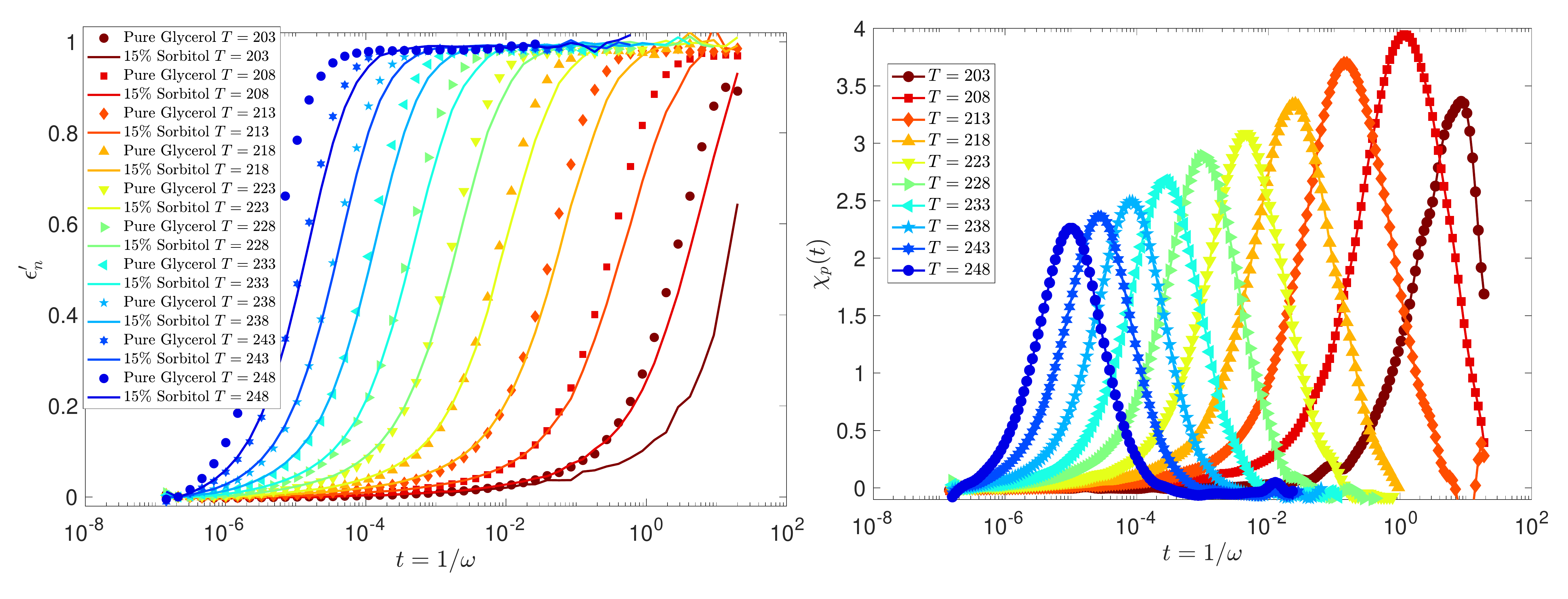}
\caption{ Left panels: Normalized storage modulus for pure Glycerol (symbols) and Glycerol-Sorbitol mixture (line) are shown for various
concentrations of Sorbitol  $c = 7.5\%$ (top panel), $10\%$ (middle panel) and $15\%$ (bottom panel) and the corresponding
Pinning Susceptibility, $\chi_p(t)$ for these Sorbitol concentrations are shown in the respective right panels. }
\label{chiP}
\end{center}
\end{figure}

In Fig.\ref{chiP} we have shown the pinning susceptibility, $\chi_p \equiv \frac{\partial \epsilon^{\prime}_n}{\partial c}$ with $c$ being
the concentration of Sorbitol in supercooled Glycerol. We have shown the plots of pinning susceptibility for $c = 7.5\%, 10\%$ and
$15\%$ ($c=5\%$ is given in the main article) as pinning susceptibility for smaller concentration ($c<5\%$) of Sorbitol is noisy
for reliable estimation of it.  

\section{Consistency check}
\begin{figure}[h!]
\begin{center}
\includegraphics[width=0.94\textwidth]{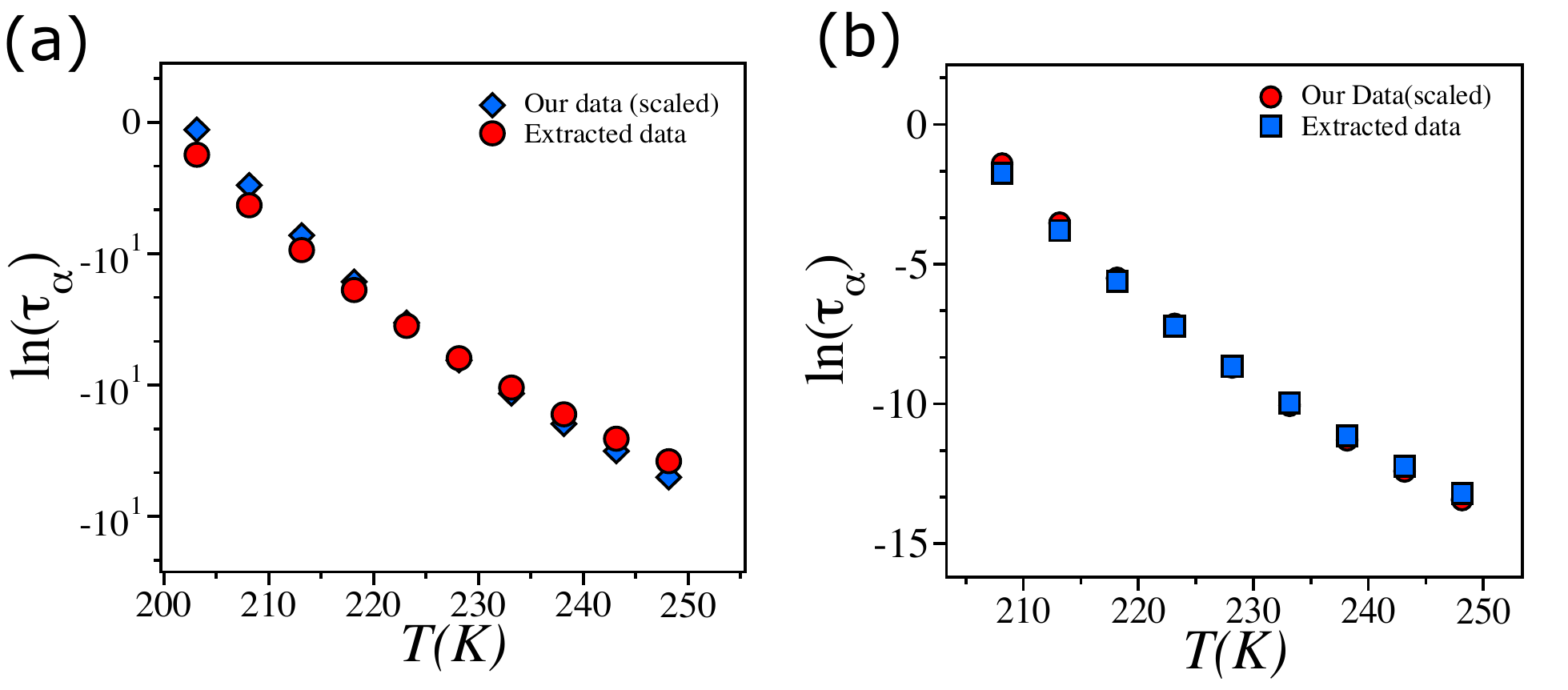}
\caption{Comparison of $\tau_\alpha$ for (a)-- pure Glycerol  and (b)-- for $c = 0.100$ of Glycerol-Sorbitol mixture. }
\label{compareTau}
\end{center}
\end{figure}

To check whether our experimental data is consistent with the previously published data, we first compare the relaxation time $\tau_\alpha$ for 
pure Glycerol ($c = 0.00$) and one fraction of the Glycerol-Sorbitol mixture ($c =0.100$(our) and $0.110$(extracted)). Apart from an overall scaling factor, our data matches well with the previously published experimental data. In Fig.~\ref{compareTau} we show the comparison.

\begin{figure}[h]
\begin{center}
\includegraphics[width=0.95\textwidth]{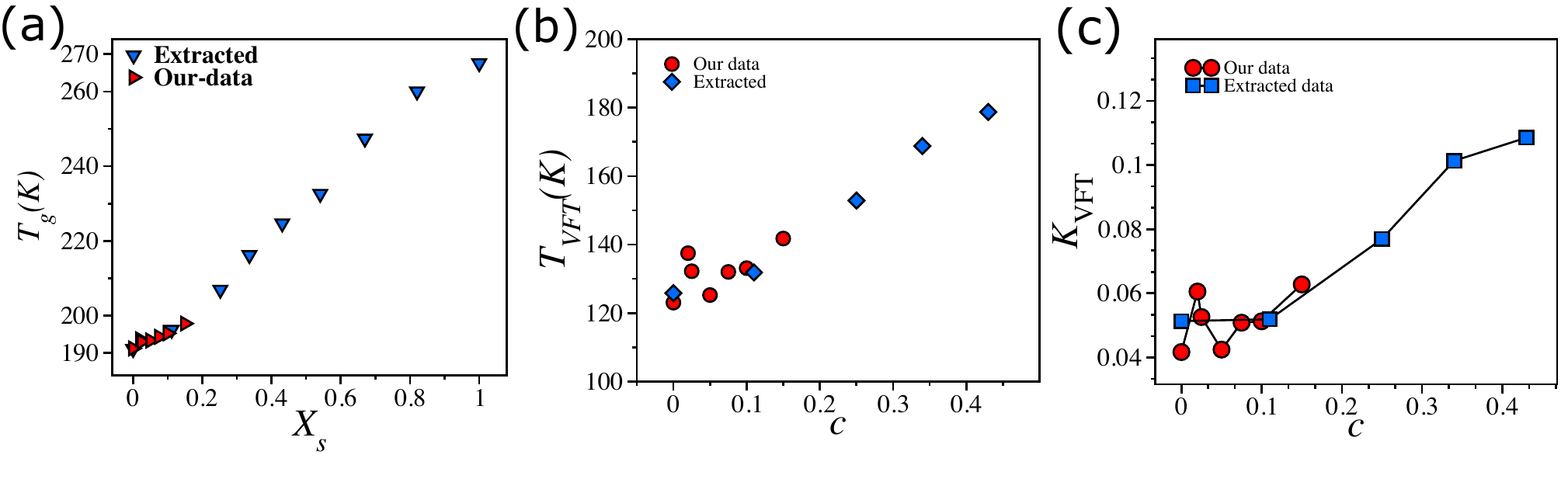}
\caption{Comparison of  (a)-- $T_g$ (scaled), (b)-- $T_{VFT}$ and (c)--kinetic fragility $K_{VFT}$ between our data and the previously published data. }
\label{consistencyCheck}
\end{center}
\end{figure}

We then compare other important quantities such as, glass transition temperature $T_g$, the $VFT$ divergence temperature $T_{VFT}$ and 
the kinetic fragility $K_{VFT}$ for our experimental data and the published data. The glass transition temperature $T_g$ is defined as 
$\tau_\alpha(T_g) = 100 s$. $T_{VFT}$ is extracted as a fitting parameter using Eq.~\eqref{VFT} to fit $T$ vs. $\tau_\alpha$ data
(See Fig.~\ref{PROCEDURE1}). The kinetic fragility defined as $K_{VFT} = \tfrac{1}{A}$ in the Eq.~\eqref{VFT} also obtained as a fitting 
parameter. In Fig.~\ref{consistencyCheck} we show such a plot for the comparison of various quantities. Note that apart from an overall 
scaling factor the $T_g$ values matches nicely with the reported data. The divergence temperature $T_{VFT}$ (no scaling factor) and the 
kinetic fragility $K_{VFT}$ also matches nicely. 

\bibliography{si-sciadvfile} 
\bibliographystyle{ScienceAdvances}